\documentclass[12pt,a4paper,openany]{book}

\newcommand{\ThesisAuthorFirstName}{Louis}

\newcommand{\ThesisAuthorFamilyName}{Owie}

\newcommand{\ThesisTitle}{FINALLY: A Dataset Recommender System for Recommender-Systems Research}

\newcommand{\ThesisResearchGroupURL}{https://isg.beel.org}

\usepackage[T1]{fontenc}

\usepackage[utf8]{inputenc}

\usepackage[english]{babel}% edited Contents -> Table of Contents
\addto\captionsenglish{%
}
\usepackage{lmodern}

\usepackage{microtype}

\usepackage[
    paper=a4paper,
    top=3cm,
    bottom=3cm,
    left=2.5cm,
    right=2.5cm,
    bindingoffset=1cm
]{geometry}

\usepackage{etoolbox}

\usepackage{blindtext}

\usepackage{enumitem}

\usepackage{csquotes}

\usepackage{xcolor}

\usepackage{graphicx}

\graphicspath{
    {Figures/}
    {Logos/}
}

\usepackage{float}

\usepackage{caption}

\usepackage{subcaption}

\usepackage{booktabs}

\usepackage{tabularx}

\usepackage{longtable}

\usepackage{array}

\usepackage{siunitx}

\usepackage{tikz}

\usepackage{pgfplots}

\pgfplotsset{
    compat=1.18
}

\usepackage{amsmath}

\usepackage{amssymb}

\usepackage{mathtools}

\usepackage{listings}

\definecolor{codegray}{gray}{0.95}

\lstdefinestyle{thesiscode}{
    backgroundcolor=\color{codegray},
    basicstyle=\ttfamily\small,
    breaklines=true,
    captionpos=b,
    frame=single,
    numbers=left,
    numberstyle=\tiny,
    showstringspaces=false,
    tabsize=4
}

\usepackage[numbers]{natbib}

\usepackage{url}

\usepackage{hyperref}

\hypersetup{
    colorlinks=true,
    linkcolor=black,
    citecolor=black,
    urlcolor=black, %edited blue -> black
    pdfborder={0 0 0},
    pdftitle={\ThesisTitle},
    pdfauthor={
        \ThesisAuthorFirstName\space
        \ThesisAuthorFamilyName
    }
}

\usepackage[nameinlink,noabbrev]{cleveref}
\usepackage[acronym,toc]{glossaries}

\makeglossaries

\newacronym{aps}{APS}{Algorithm Performance Space}

\newacronym{api}{API}{Application Programming Interface}

\newacronym{ndcg}{NDCG}{Normalized Discounted Cumulative Gain}

\newacronym{effcov}{EffCov}{Effective Covariance}

\newacronym{hr}{HR}{Hit Rate}

\newacronym{html}{HTML}{Hypertext Markup Language}

\newacronym{json}{JSON}{JavaScript Object Notation}

\newacronym{mnnd}{MNND}{Mean Nearest Neighbor Distance}

\newacronym{pca}{PCA}{Principal Component Analysis}

\newacronym{png}{PNG}{Portable Network Graphics}

\newacronym{url}{URL}{Uniform Resource Locator}

\usepackage{pdfpages}

\usepackage{emptypage}
\usepackage{doi}

\makeatletter
\renewcommand{\cleardoublepage}{%
    \clearpage
    \if@twoside
        \ifodd\c@page
        \else
            \stepcounter{page}%
        \fi
    \fi
}
\makeatother

\begin{document}

% ============================================================================
% TITLE PAGE
%
% All values shown here are defined in the configuration block at the top.
% ============================================================================

% Public/arXiv title page
\thispagestyle{empty}
\includepdf[
    pages=1,
    pagecommand={},
    fitpaper=true
]{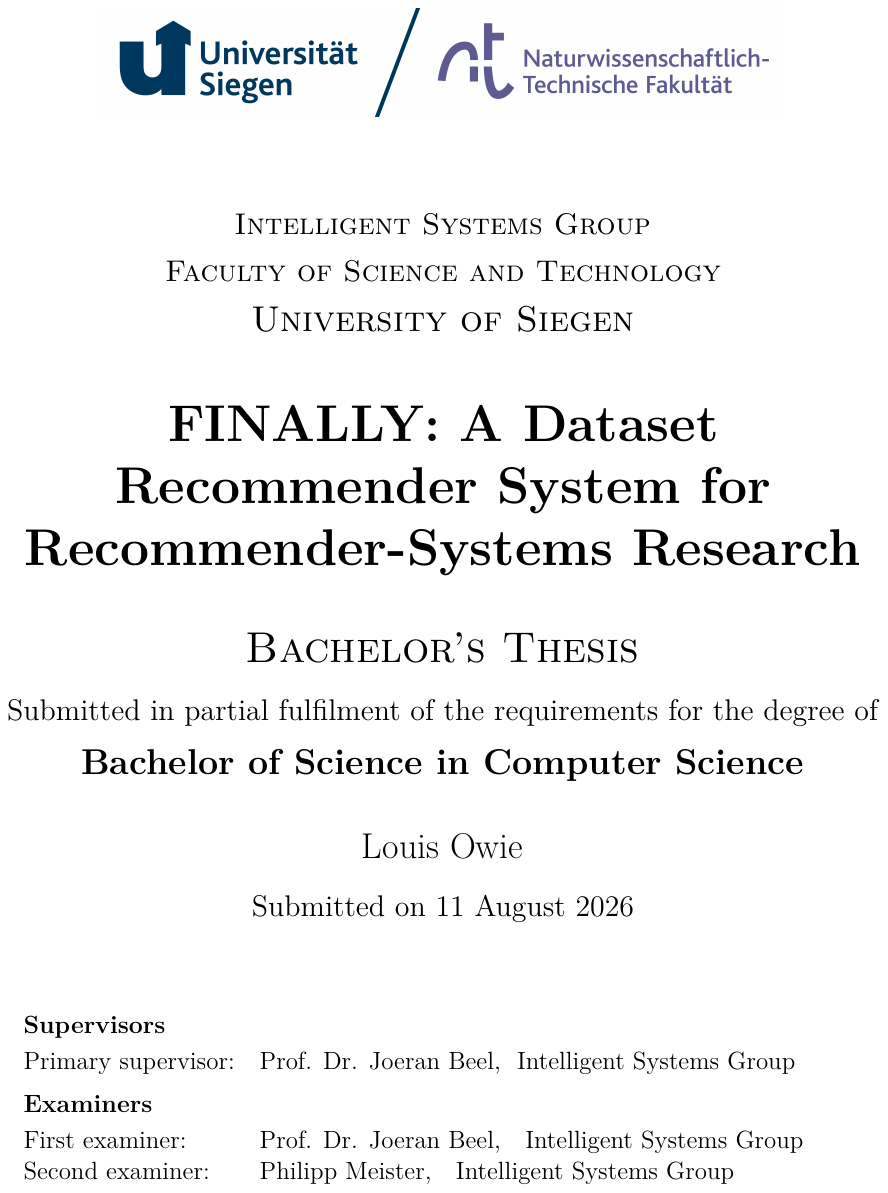}

% ============================================================================
% FRONT MATTER
%
% Roman page numbers are used automatically by \frontmatter.
% ============================================================================

\thispagestyle{empty} %edited

\frontmatter
\pagenumbering{Roman} %edited

\pagestyle{plain}

% ----------------------------------------------------------------------------
% ACKNOWLEDGEMENTS
%
% The acknowledgements deliberately appear at the beginning of the thesis.
%
% Mention, where applicable:
%
% - academic supervision,
% - technical support,
% - funding,
% - access to data or infrastructure,
% - participants or external collaborators, and
% - persons who contributed to the work but not sufficiently for authorship.
%
% Follow the applicable university rules concerning the disclosure of
% generative AI and other electronic tools.
% ----------------------------------------------------------------------------

\cleardoublepage

\phantomsection

\addcontentsline{toc}{chapter}{Acknowledgements}

\chapter*{Acknowledgements}

I would like to thank Prof. Dr. Joeran Beel for supervising this thesis and for his constructive feedback throughout the development of FINALLY. I would also like to thank Tobias Vente for his helpful suggestions during the early stages of the project and for his support in preparing the related demo paper. I am very grateful to both of them for the time and effort they invested in supporting this work.

% ----------------------------------------------------------------------------
% ABSTRACT
%
% Summarise the following elements in one coherent text:
%
% 1. the specific background,
% 2. the research problem,
% 3. the objective,
% 4. the methodology,
% 5. the principal results, and
% 6. the main conclusion or contribution.
%
% Focus on the actual findings. Report concrete numbers for the central
% results instead of writing only that one method performed better.
%
% Avoid spending most of the abstract on broad background information.
% Citations should normally not be necessary in the abstract.
% ----------------------------------------------------------------------------

\cleardoublepage

\phantomsection

\addcontentsline{toc}{chapter}{Abstract}

\chapter*{Abstract}

Dataset selection shapes the empirical conditions under which recommender-system algorithms are evaluated, yet existing tools provide limited support for constructing complete dataset sets that jointly satisfy experimental constraints and set-level selection objectives. To address this problem, I developed FINALLY, a web-based dataset recommender for constructing configurable dataset sets for offline recommender-systems evaluations. FINALLY combines required datasets, candidate-pool restrictions, metadata filters, configurable target-set sizes, Random selection, and diverse and non-diverse strategies based on adapted Effective Covariance and Convex Hull objectives.

I evaluated FINALLY through 420 recommendation runs across ten systematically varied configurations. All evaluated dataset sets satisfied the applicable target-size, duplicate-avoidance, snapshot-membership, required-dataset, and metadata-filter requirements. All 40 deterministic strategy--configuration combinations were reproducible. Both the Effective-Covariance-based and Convex-Hull-based strategies produced the expected diverse-versus-non-diverse score ordering in all ten configurations. Under their corresponding objectives, the diverse strategies produced scores above all 30 configuration-specific Random results, whereas the non-diverse strategies produced scores below all 30 Random results.

These results establish technical consistency for the evaluated FINALLY workflow and show that the implemented strategies follow their intended optimization directions within the investigated configuration space. They do not establish the scientific suitability, global optimality, or practical superiority of the generated selections.

% ----------------------------------------------------------------------------
% TABLE OF CONTENTS
% ----------------------------------------------------------------------------

\cleardoublepage

\phantomsection

\addcontentsline{toc}{chapter}{Table of Contents}

\tableofcontents

% ----------------------------------------------------------------------------
% LIST OF FIGURES
% ----------------------------------------------------------------------------

\cleardoublepage

\phantomsection

\addcontentsline{toc}{chapter}{List of Figures}

\listoffigures

% ----------------------------------------------------------------------------
% LIST OF TABLES
% ----------------------------------------------------------------------------

\cleardoublepage

\phantomsection

\addcontentsline{toc}{chapter}{List of Tables}

\listoftables

% ----------------------------------------------------------------------------
% LIST OF SOURCE-CODE LISTINGS
% ----------------------------------------------------------------------------

\cleardoublepage

%\phantomsection

%\addcontentsline{toc}{chapter}{List of Listings}

%\lstlistoflistings

%\clearpage %edited
% ----------------------------------------------------------------------------
% ACRONYM LIST
%
% \glsaddall ensures that every defined acronym appears in the list.
% Remove \glsaddall if only acronyms actually used in the thesis should appear.
% ----------------------------------------------------------------------------

\thispagestyle{empty} %edited
% \glsaddall

\printglossary[
    type=\acronymtype,
    title=Acronyms
]

% Reset acronym usage so that acronyms are expanded again in the main text.
\glsresetall

% ============================================================================
% MAIN MATTER
%
% Arabic page numbers and numbered chapters start here.
% ============================================================================

\mainmatter

% ============================================================================
% CHAPTER 1: INTRODUCTION
%
% Keep the introductory background specific to this thesis. A background
% section that could be copied into almost any computer-science thesis is too
% broad.
%
% State the research problem explicitly and quantify it where possible.
% Derive the research questions from that problem.
%
% Distinguish:
%
% - research problem,
% - research question,
% - research objective,
% - research tasks, and
% - research contribution.
%
% Implementing software, conducting a survey, or running an evaluation are
% tasks. The resulting method, dataset, software, or new knowledge may
% constitute a contribution.
% ============================================================================

\chapter{Introduction}
\label{chap:introduction}

This chapter establishes the motivation, research problem, and scope of this thesis. It derives the research questions and research goal, summarizes the main contributions, and outlines the structure of the thesis.

\section{Background and Motivation}
\label{sec:introduction-background}

Offline evaluation is widely used to compare recommender-system algorithms using previously collected interaction data \cite{bauer2024evaluation,gunawardana2022evaluating}. It enables researchers to evaluate multiple algorithms and metrics under controlled experimental conditions without deploying the systems in a live environment. However, the resulting evidence depends on the specific evaluation design. Decisions concerning the data, experimental protocol, and evaluation metrics can substantially influence algorithm comparisons and the conclusions drawn from them \cite{canamares2020offline}.

The selected datasets constitute a central part of this experimental design. They determine the users, items, interactions, domains, and data characteristics under which algorithms are evaluated. Prior research has shown that characteristics of rating data can systematically affect recommendation performance \cite{adomavicius2012data}. Comparisons across multiple datasets further demonstrate that algorithms can exhibit different relative performance patterns depending on the dataset used \cite{beel2024aps}. A dataset is therefore not merely an interchangeable input to an experiment. It defines an empirical condition under which claims about algorithm behavior are examined.

Dataset selection should consequently be aligned with the research question and the intended scope of an evaluation. Researchers may need datasets that satisfy practical constraints, represent different empirical conditions, or complement datasets that have already been used. Constructing such a dataset set requires more than identifying individually available or popular datasets. It requires an informed selection process that considers the complete set and the objective it is intended to serve. However, current tools provide only partial support for constructing such dataset sets.

\section{Research Problem}
\label{sec:research-problem}

Despite the methodological importance of dataset selection, the reasoning behind this decision is frequently not made explicit in recommender-systems research. An analysis of all 58 full papers published at the ACM Conference on Recommender Systems in 2024 found that only 8 papers, corresponding to 14\%, explained why their selected datasets were suitable for the respective experiment. The remaining 50 papers, corresponding to 86\%, did not provide such a justification \cite{vente2025apsexplorer}. The same analysis found that dataset usage was concentrated on several frequently used dataset families: Amazon datasets were used in 22 of the examined papers, MovieLens in 20, Yelp in 9, and Gowalla in 7 \cite{vente2025apsexplorer}.

The absence of an explicit justification does not imply that the selected datasets were unsuitable. Researchers may have had valid reasons that were not reported in the publication. However, missing justifications make it difficult to assess whether the selected datasets match the experimental objective and whether relevant empirical conditions are represented. They also reduce the transparency of the evaluation design and complicate the interpretation of how far the resulting conclusions may extend beyond the specific datasets used.

This concentration on frequently used datasets raises a related concern. Popular datasets provide recognizable benchmarks and facilitate comparisons with prior work, but popularity alone does not establish their suitability for a particular research question. Moreover, selecting several frequently used datasets does not necessarily create a varied experimental setting because datasets can exhibit similar algorithm-performance patterns despite differences in their names, domains, or metadata \cite{beel2024aps}. Conversely, selecting datasets solely because they appear different does not guarantee that they satisfy the requirements of the experiment. An informed selection must therefore consider both the properties of the individual datasets and the objective of the complete dataset set.

Existing systems provide partial support for dataset selection. General-purpose dataset search engines help researchers discover datasets distributed across online repositories \cite{brickley2019googledatasetsearch}. Dataset recommendation approaches such as DataFinder retrieve individually relevant datasets from descriptions of research ideas \cite{viswanathan2023datafinder}. Within recommender-systems research, the APS Explorer supports interactive inspection of dataset metadata and algorithm-performance patterns in an \gls{aps} \cite{vente2025apsexplorer}. These systems facilitate the discovery, recommendation, or comparison of individual datasets. However, researchers must still assess the available options and manually assemble the complete dataset set used in an evaluation.

The need to construct complete dataset sets occurs in several experimental situations. Researchers may need to create an initial dataset set when no datasets have been selected, extend an existing selection while retaining datasets that are already required, or choose additional datasets for post-development validation. In each case, the final selection may have to satisfy dataset-level constraints, reach a specified target-set size, include previously selected datasets, and reflect a particular set-level selection objective. These requirements concern the dataset set as a whole and cannot be addressed solely by ranking datasets according to their individual relevance.

The research problem addressed in this thesis is therefore the lack of an integrated, recommendation-centered workflow for constructing configurable dataset sets for offline recommender-systems evaluations. Such a workflow should combine required datasets, dataset-level constraints, a configurable target-set size, and alternative dataset-set selection objectives. It should also allow researchers to inspect the resulting recommendation while preserving their responsibility for determining whether the selected datasets are appropriate for the intended experiment. The system is thus intended to provide decision support rather than to establish dataset suitability automatically.

Based on this research problem, the thesis addresses the following research
questions:

\begin{description}
    \item[RQ1:] How can the construction of configurable dataset sets for offline recommender-systems evaluations be supported through an integrated, recommendation-centered workflow?

    \item[RQ2:] To what extent do the generated dataset sets satisfy user-defined constraints and reflect the selected recommendation strategy across different configurations?
\end{description}

\section{Research Goal}
\label{sec:research-goal}

The goal of this thesis is to develop an integrated, recommendation-centered system for constructing configurable dataset sets for offline recommender-systems evaluations and to evaluate whether its recommendations satisfy user-defined constraints and reflect the selected strategy across different configurations.

To pursue this goal, I developed FINALLY, a web-based dataset recommender for constructing configurable dataset sets, and systematically evaluated its recommendations across varied configurations. FINALLY is designed to combine required datasets, dataset-level constraints, target-set sizes, and alternative recommendation strategies within a single workflow. It is intended to support dataset-selection decisions rather than determine whether a dataset set is scientifically suitable for a particular experiment. The final assessment therefore remains the responsibility of the researcher.

\section{Contributions}
\label{sec:contributions}

This thesis makes three main contributions:

\begin{enumerate}
    \item \textbf{FINALLY dataset recommender.} I developed FINALLY, an open-source, web-based dataset recommender that supports the construction, inspection, and documentation of configurable dataset sets for offline recommender-systems evaluations. The system combines required datasets, dataset-level constraints, configurable target-set sizes, and alternative recommendation strategies within an integrated workflow. FINALLY builds on the data infrastructure and code base of APS Explorer \cite{vente2025apsexplorer}. The contribution of this thesis is the recommendation-centered workflow and the dataset-recommender functionality developed around it rather than the original APS Explorer functionality.

    \item \textbf{Operational integration of dataset-set selection strategies.} I integrated Random selection and adapted implementations of the \gls{aps}-based \gls{effcov} and Convex Hull diversity measures described by Reising~\cite{reising2026systematic} into FINALLY's configurable recommendation workflow. Both \gls{aps}-based objectives are provided in diverse and non-diverse variants and are combined with required datasets, metadata filters, performance metrics, cutoffs, and target-set sizes. The underlying \gls{aps} concepts and diversity measures are not contributions of this thesis. My contribution lies in adapting and integrating these measures into an executable dataset-set recommender.

    \item \textbf{Empirical characterization of technical consistency and strategy behavior.} I conducted a systematic technical and algorithmic evaluation of FINALLY across 420 recommendation runs. The results show that all evaluated dataset sets satisfied the configured target-size, snapshot-membership, required-dataset, duplicate-avoidance, and metadata-filter requirements and that all 40 deterministic strategy--configuration combinations were reproducible. Both \gls{effcov} and Convex Hull showed the expected strategy alignment across all ten investigated configurations, with the corresponding diverse and non-diverse strategies positioned on opposite sides of the configuration-specific Random reference distributions. The evaluation additionally characterizes end-to-end response times across recommendation strategies and target-set sizes.
\end{enumerate}

\section{Structure of the Thesis}
\label{sec:thesis-structure}

The remainder of this thesis is structured as follows. Chapter~\ref{chap:related-work} introduces the conceptual foundations of performance-based dataset selection, reviews existing approaches to dataset discovery, recommendation, selection, and exploration, and positions the research problem addressed by this thesis within the existing literature.

Chapter~\ref{chap:methodology} describes the research design, requirement derivation, iterative system development, and verification process. Chapter~\ref{chap:finally} presents FINALLY, including its architecture, recommendation workflow, dataset-set selection procedures, and result-inspection functionality. Chapter~\ref{chap:evaluation} defines the systematic evaluation setup, while Chapter~\ref{chap:results} reports the resulting measurements and observations. Chapter~\ref{chap:discussion} interprets these findings, examines their practical implications and threats to validity, and derives explicit answers to the research questions. Finally, Chapter~\ref{chap:conclusion} synthesizes the main findings and conclusions, and Chapter~\ref{chap:future-work-limitations} discusses the limitations of the thesis and directions for future work.

% ============================================================================
% NEW CHAPTER 2: RELATED WORK
% ============================================================================
\chapter{Related Work}
\label{chap:related-work}

This chapter reviews the research relevant to constructing dataset sets for offline recommender-systems evaluations. Section~\ref{sec:related-work-practices} examines findings on dataset-selection practices and their methodological consequences. Section~\ref{sec:related-work-search} reviews approaches to dataset search and recommendation. Section~\ref{sec:related-work-aps} then introduces performance-based dataset representations, the set-level diversity measures on which FINALLY's selection objectives are based, and existing \gls{aps}-based approaches to dataset selection and exploration. Finally, Section~\ref{sec:related-work-synthesis} synthesizes these research strands and derives the research gap addressed by FINALLY.

\section{Dataset-Selection Practices in Recommender-Systems Research}
\label{sec:related-work-practices}

Empirical research on recommender-systems evaluation identifies dataset selection as an influential part of experimental design. Bauer et al. document the variety of methodological decisions involved in recommender-systems evaluation and the continued prevalence of offline experiments \cite{bauer2024evaluation}. Analyses of dataset usage additionally show that recommender-systems research concentrates on a limited number of established benchmarks whose characteristics and suitability vary substantially \cite{chin2022datasets,beel2024aps}. These findings motivate treating dataset selection as an experimental decision that requires consideration of the intended research setting rather than as a neutral preliminary step.

The characteristics and origin of individual datasets can affect how experimental results should be interpreted. Chin et al. identify substantial differences between commonly used recommendation datasets \cite{chin2022datasets}, while Fan et al. show for MovieLens that recorded interactions reflect the platform, available items, and recommendation mechanisms involved in collecting the data \cite{fan2024movielens}. The concrete dataset instance may additionally depend on preprocessing: Beel and Brunel report that recommender-systems researchers frequently use pruned versions of available datasets \cite{beel2019pruning}. Variations in preprocessing, splitting, and evaluation protocols can consequently cause identically named benchmarks to represent different experimental conditions and can affect reproducibility and algorithm comparisons \cite{beel2016reproducibility,sun2020evaluating}.

The composition of a multi-dataset evaluation introduces a related set-level decision. Shevchenko et al. evaluate 11 collaborative-filtering algorithms across 30 datasets and show that dataset variability affects algorithm comparisons and aggregated rankings \cite{shevchenko2024variability}. Evaluating several datasets can therefore reduce dependence on a single benchmark, but the number of datasets alone does not establish that the resulting set represents complementary empirical conditions. Several datasets may still induce similar algorithm behavior and provide partly redundant evidence.

Existing work documents these problems and provides methodological recommendations for dataset selection and evaluation \cite{bauer2024evaluation,chin2022datasets,shevchenko2024variability}. However, it does not provide an integrated workflow for constructing a complete dataset set under combined dataset-level and set-level requirements. The following sections review technical approaches that address parts of this process through dataset search, recommendation, performance-based exploration, and dataset-set optimization.

\section{Dataset Search and Recommendation}
\label{sec:related-work-search}

Dataset search addresses the problem of matching an information need with potentially suitable datasets. Chapman et al. characterize dataset search as the discovery, exploration, and return of datasets to an end user \cite{chapman2020datasetsearch}. Existing systems typically return ranked datasets accompanied by metadata, previews, explanations, or filtering functionality. Their primary task is therefore the discovery and assessment of individual datasets rather than the joint construction of an evaluation set.

Google Dataset Search applies metadata-based dataset retrieval at web scale. The system discovers structured metadata published by dataset providers, reconciles descriptions referring to the same dataset, and links users to the repositories hosting the data \cite{brickley2019googledatasetsearch}. This improves access to datasets distributed across repositories and organizational websites, but its retrieval objective ranks individual datasets and does not consider relationships among several selected results.

Scientific dataset recommendation extends conventional search by accepting descriptions of the intended research task. DataFinder retrieves datasets from natural-language descriptions of research ideas using a trained bi-encoder \cite{viswanathan2023datafinder}. Although this enables researchers to express requirements concerning domain, modality, or scale more naturally than through keyword matching, DataFinder still ranks individual datasets rather than optimizing the composition of a complete dataset set.

Domain-specific approaches extend this retrieval setting further. The DS4RS preprint describes a search engine for recommender-systems datasets that performs semantic search over dataset names, descriptions, and recommendation domains and provides explanations for individual results \cite{shao2025ds4rs}. TeG-DRec combines textual descriptions with graph-based relationships for scientific dataset recommendation \cite{qayyum2025tegdrec}. Despite their different representations, both approaches retain the fundamental objective of ranking individually relevant datasets.

The reviewed systems therefore improve dataset discovery through metadata, natural-language descriptions, domain-specific information, or graph-based relationships. Their objectives do not, however, determine whether several results provide complementary algorithm-performance conditions, allow previously selected datasets to remain fixed, or optimize a complete selection for a configurable target-set size. Dataset discovery and coordinated dataset-set construction consequently represent related but distinct problems.

\section{Performance-Based Dataset Selection}
\label{sec:related-work-aps}

\subsection{Algorithm Performance Spaces}
\label{sec:related-work-aps-foundations}

Performance-based representations characterize instances through observed algorithm behavior rather than exclusively through descriptive features. Related approaches include Instance Space Analysis, which combines instance characteristics and algorithm-performance information to study benchmark coverage and regions of algorithm strength and weakness \cite{smithmiles2023instance}, and Algorithm-Performance Personas, which group instances according to similarities in algorithm behavior \cite{tyrrell2020personas}. Beel et al. transfer this performance-oriented perspective to complete recommender-systems datasets through \glspl{aps} \cite{beel2024aps}.

Let \(A=\{a_1,\ldots,a_n\}\) denote a portfolio of recommendation algorithms and let \(p_{m,k}(a_i,d)\) denote the performance of algorithm \(a_i\) on dataset \(d\), measured using metric \(m\) at cutoff \(k\). A dataset is represented in the corresponding \gls{aps} by the performance vector

\begin{equation}
    \mathbf{x}_{d}^{(m,k)}
    =
    \bigl(
        p_{m,k}(a_1,d),
        \ldots,
        p_{m,k}(a_n,d)
    \bigr).
    \label{eq:aps-performance-vector}
\end{equation}

Each algorithm therefore defines one dimension and each dataset one point in the performance space \cite{beel2024aps,vente2025apsexplorer}. The resulting representation is conditional on the algorithm portfolio, performance metric, cutoff, and underlying evaluation results.

Distances between dataset points describe differences in algorithm-performance patterns. Nearby datasets exhibit similar performance values across the represented algorithms, whereas larger distances indicate more distinct performance patterns \cite{beel2024aps}. \gls{aps}-based diversity therefore refers to variation in algorithm behavior across datasets rather than to differences between the algorithms themselves \cite{beel2024aps,reising2026systematic}. This notion differs from metadata-based similarity: datasets with similar domains, sizes, or densities may induce different algorithm behavior, while descriptively different datasets may exhibit similar performance patterns. Metadata and \gls{aps}-based representations therefore capture complementary aspects of dataset similarity.

Because an \gls{aps} contains one dimension per represented algorithm, visual exploration commonly requires dimensionality reduction. Beel et al. use \gls{pca} to project the high-dimensional performance vectors into two dimensions \cite{beel2024aps}. Such a projection supports visualization but does not preserve every relationship of the original space, and its axes no longer correspond directly to individual algorithms \cite{jolliffe2016pca}. Consequently, the two-dimensional representation is suitable for inspection, whereas the set-level objectives discussed below operate on the original high-dimensional performance vectors.

\subsection[Dataset-Set Diversity in APS]{Dataset-Set Diversity in \gls{aps}}
\label{sec:related-work-diversity}

Selecting several datasets in an \gls{aps} produces a set of points whose geometric arrangement reflects variation in algorithm-performance patterns. Reising proposes two set-level objectives for quantifying this arrangement: a Convex-Hull-based objective and a covariance-based objective referred to as \gls{effcov} \cite{reising2026systematic}. Both combine local separation between datasets with a measure of the broader geometric structure of the complete set.

Let \(S=\{\mathbf{x}_1,\ldots,\mathbf{x}_n\}\) denote a set of \(n\geq2\) dataset-performance vectors. Local separation is quantified through the \gls{mnnd},

\begin{equation}
    \operatorname{MNND}(S)
    =
    \frac{1}{n}
    \sum_{i=1}^{n}
    \min_{\substack{j \in \{1,\ldots,n\}\\j\neq i}}
    \left\lVert \mathbf{x}_i-\mathbf{x}_j \right\rVert_2.
    \label{eq:mean-nearest-neighbor}
\end{equation}

A larger \gls{mnnd} indicates stronger average separation from the nearest selected alternative. Because this local measure does not describe the global arrangement of the point set, Reising combines it with an additional structure term \cite{reising2026systematic}.

\paragraph{Convex-Hull-based diversity.}

The convex hull of a point set is the smallest convex set containing all of its points \cite{barber1996quickhull}. Its volume describes the geometric extent covered by the outermost vectors. In a high-dimensional \gls{aps}, however, \(n\) selected points span at most an \((n-1)\)-dimensional affine subspace. Reising therefore calculates the hull in the intrinsic affine subspace of the selected vectors \cite{reising2026systematic}.

Let \(X_S\) denote the centered matrix of the selected vectors and \(r=\operatorname{rank}(X_S)\). After projection onto an orthonormal basis of the \(r\)-dimensional affine span, the intrinsic structure term is

\begin{equation}
    \operatorname{CH}_{\mathrm{intr}}(S)
    =
    \operatorname{Vol}_{r}
    \left(
        \operatorname{conv}\left(P_r(S)\right)
    \right),
    \label{eq:intrinsic-convex-hull}
\end{equation}

where \(P_r(S)\) denotes the projected point set. For \(r=1\), the volume is the length of the spanned interval; for \(r=2\), it is the corresponding area; and higher ranks produce the respective hypervolume. If the selected vectors have intrinsic rank \(r=0\), the structure term is defined as zero.

Reising's complete Convex-Hull-based diversity metric is

\begin{equation}
    D^{\mathrm{CH}}(S)
    =
    \operatorname{MNND}(S)
    \cdot
    \left(
        \operatorname{CH}_{\mathrm{intr}}(S)
        +
        \varepsilon
    \right),
    \label{eq:convex-hull-diversity}
\end{equation}

where \(\varepsilon>0\) is a stabilizing constant \cite{reising2026systematic}. The \gls{mnnd} term captures local separation, while the intrinsic hull volume represents the geometric extent of the complete set. The stabilizer prevents a zero structure term from eliminating the spacing contribution entirely.

Because Convex Hull volume is determined by the boundary of a point set, vectors inside an existing hull do not increase its volume. The measure additionally depends on the estimation of the intrinsic rank and the construction of a potentially high-dimensional hull. It should therefore be interpreted as spread within the subspace spanned by the selected datasets rather than as strict coverage of the complete \gls{aps}.

\paragraph{Effective-Covariance-based diversity.}

\gls{effcov} replaces the hull volume with a covariance-based structure term. Let \(\mathbf{p}_1,\ldots,\mathbf{p}_n \in \mathbb{R}^{d}\) denote the selected performance vectors. Their mean is

\begin{equation}
    \bar{\mathbf{p}}
    =
    \frac{1}{n}
    \sum_{i=1}^{n}
    \mathbf{p}_i,
\end{equation}

and the centered performance matrix \(X\in\mathbb{R}^{n\times d}\) is

\begin{equation}
    X
    =
    \begin{bmatrix}
        (\mathbf{p}_1-\bar{\mathbf{p}})^{\mathsf{T}} \\
        \vdots \\
        (\mathbf{p}_n-\bar{\mathbf{p}})^{\mathsf{T}}
    \end{bmatrix}.
\end{equation}

The sample covariance matrix is

\begin{equation}
    C
    =
    \frac{1}{n-1}
    X^{\mathsf{T}}X.
\end{equation}

Its eigenvalues describe variance along orthogonal directions of the centered performance space \cite{reising2026systematic,jolliffe2016pca}. Let \(d_{\mathrm{eff}}\) denote the number of retained positive covariance eigenvalues, with \(d_{\mathrm{eff}}\leq\min(d,n-1)\), and let \(\lambda_1,\ldots,\lambda_{d_{\mathrm{eff}}}\) denote these eigenvalues. For \(d_{\mathrm{eff}}>0\), the structure term is

\begin{equation}
    \operatorname{EffCov}_{\mathrm{struct}}(S)
    =
    \left(
        \prod_{i=1}^{d_{\mathrm{eff}}}
        \lambda_i
    \right)^{\frac{1}{2d_{\mathrm{eff}}}},
    \label{eq:effective-covariance}
\end{equation}

whereas \(\operatorname{EffCov}_{\mathrm{struct}}(S)=0\) if \(d_{\mathrm{eff}}=0\). For positive \(d_{\mathrm{eff}}\), the expression corresponds to the geometric mean of the standard deviations along the retained principal directions.

The complete \gls{effcov} diversity metric is

\begin{equation}
    D^{\mathrm{EffCov}}(S)
    =
    \operatorname{MNND}(S)
    \cdot
    \left(
        \operatorname{EffCov}_{\mathrm{struct}}(S)
        +
        \varepsilon
    \right).
    \label{eq:effcov-diversity}
\end{equation}

Unlike convex-hull volume, the covariance-based term depends on all selected vectors rather than only on points forming the boundary. However, \gls{effcov} does not explicitly reward the number of effective dimensions. A selection with strong variation in comparatively few directions may therefore obtain a higher score than a more compact selection spanning additional directions \cite{reising2026systematic}.

The raw values of both measures depend on the scale and structure of the underlying \gls{aps} and on the selected set size. Reising therefore additionally defines the normalized score

\begin{equation}
    \widehat{D}(S)
    =
    \frac{D(S)}{D^*(k)},
    \label{eq:normalized-aps-diversity}
\end{equation}

where \(D^*(k)\) denotes the maximum attainable value among all subsets of size \(k\) in the respective \gls{aps} \cite{reising2026systematic}. Provided that \(D^*(k)>0\), this expresses diversity relative to the most diverse possible set under the same conditions.

A diversity-oriented selection strategy maximizes the corresponding Convex Hull or \gls{effcov} score, whereas minimizing the same score favors more compact algorithm-performance patterns. These objectives define diversity relative to the selected \gls{aps}; they do not independently establish whether a dataset set is scientifically suitable for a particular experiment.

\subsection[APS-Based Selection and Exploration]{\gls{aps}-Based Selection and Exploration}
\label{sec:related-work-aps-approaches}

The introduction of \gls{aps} enables dataset selection to consider algorithm-performance patterns in addition to descriptive metadata. Subsequent work extends this representation toward quantitative selection criteria, interactive exploration, and explicit dataset-set optimization.

Schulz proposes strategic \gls{aps}-based criteria for dataset difficulty, performance variance, and diversity and evaluates their use for testing assumptions about datasets and identifying strategically different selections \cite{schulz2025strategic}. This work extends visual performance-space interpretation with quantitative criteria for individual datasets and complete selections, but focuses on the formulation and experimental analysis of these measures rather than an operational recommendation workflow.

APS Explorer operationalizes interactive \gls{aps} exploration through a web application\footnote{\url{https://datasets.recommender-systems.com}} \cite{vente2025apsexplorer}. Its modules support multidimensional \gls{aps} exploration, pairwise algorithm comparison, and metadata-based dataset comparison. Researchers can filter and inspect datasets and use their performance-space positions to inform a manual selection. APS Explorer therefore connects metadata-based and performance-based inspection, but it does not automatically construct a complete dataset set according to a configurable set-level objective.

Reising directly addresses \gls{aps}-based dataset-set optimization using the Convex Hull and \gls{effcov} objectives defined in Section~\ref{sec:related-work-diversity} \cite{reising2026systematic}. A greedy multi-start procedure approximates diverse and non-diverse subsets without enumerating every possible combination. Reising evaluates the resulting selections in terms of generalization and informativeness and reports that diversity-oriented selections cover broader performance ranges and induce more varied algorithm rankings, whereas Random selection better approximates aggregate performance across the complete dataset collection \cite{reising2026systematic}.

Reising's contribution provides the most direct algorithmic basis for configurable dataset-set optimization in FINALLY. The proposed procedures are presented as a selection methodology and experimental pipeline rather than as a public end-user recommendation system. They do not combine the set-level objectives with an interface for retaining required datasets, applying dataset-level filters, configuring performance settings, inspecting the resulting selection, and exporting its configuration and result.

The \gls{aps} research line thus provides complementary components for performance-based dataset selection: Beel et al. establish the dataset representation \cite{beel2024aps}, Schulz develops quantitative strategic criteria \cite{schulz2025strategic}, APS Explorer provides interactive exploration \cite{vente2025apsexplorer}, and Reising formulates explicit dataset-set optimization \cite{reising2026systematic}.

\section{Synthesis and Research Gap}
\label{sec:related-work-synthesis}

The reviewed research addresses three complementary aspects of systematic dataset selection. Studies of evaluation practices establish that benchmark choice, dataset variation, and preprocessing can affect the conclusions drawn from recommender-systems experiments \cite{chin2022datasets,shevchenko2024variability}. Dataset-search and recommendation systems support the discovery of individually relevant datasets \cite{brickley2019googledatasetsearch,viswanathan2023datafinder}. Performance-based research provides representations, interactive exploration, and objectives for constructing dataset selections according to algorithm-performance patterns \cite{beel2024aps,reising2026systematic}.

Table~\ref{tab:related-work-comparison} compares representative dataset-search and \gls{aps}-based approaches by output granularity, use of algorithm-performance information, set-level objectives, and availability of an interactive end-user interface.

\begin{table}[htbp]
    \centering
    \small
    \renewcommand{\arraystretch}{1.12}
    \caption[Comparison of representative dataset-search and APS-based approaches.]{Comparison of representative dataset-search and \gls{aps}-based approaches.}
    \label{tab:related-work-comparison}
    \begin{tabular}{
        @{}
        p{0.28\textwidth}
        p{0.30\textwidth}
        c
        c
        c
        @{}
    }
        \toprule
        Approach & Output & Perf. & Set obj. & Interactive UI \\
        \midrule

        Google Dataset Search \cite{brickley2019googledatasetsearch}
        & Ranked datasets
        & No
        & No
        & Yes \\

        DataFinder \cite{viswanathan2023datafinder}
        & Ranked datasets
        & No
        & No
        & No \\

        DS4RS \cite{shao2025ds4rs}
        & Ranked datasets
        & No
        & No
        & Yes \\

        \addlinespace

        \gls{aps} \cite{beel2024aps}
        & Dataset representation and informed selection
        & Yes
        & No
        & No \\

        Schulz \cite{schulz2025strategic}
        & Scored datasets and selections
        & Yes
        & Yes
        & No \\

        APS Explorer \cite{vente2025apsexplorer}
        & Visual exploration and manual selection
        & Yes
        & No
        & Yes \\

        Reising \cite{reising2026systematic}
        & Optimized dataset sets
        & Yes
        & Yes
        & No \\

        \bottomrule
    \end{tabular}

    \vspace{0.4em}
    \parbox{\textwidth}{\footnotesize\textit{Notes:} Perf. indicates whether an approach uses algorithm-performance information. Set obj. indicates whether it assesses or optimizes a property of a complete dataset selection. Interactive UI indicates whether the publication describes an interactive end-user interface.}
\end{table}

Three of the seven approaches provide an interactive user interface, but these systems support dataset search or manual exploration rather than optimization of a complete dataset set. Schulz and Reising are the two approaches in the comparison that assess complete selections through set-level criteria, with Reising explicitly formulating diverse and non-diverse dataset-set construction as an optimization problem. None of the seven approaches combines set-level optimization with the complete configuration and inspection workflow considered in this thesis.

Dataset-level eligibility and set-level performance properties address complementary requirements. A dataset may be individually relevant to a research task without contributing a distinct algorithm-performance condition to the complete evaluation set. Conversely, an \gls{aps}-complementary dataset may fail study-specific requirements concerning its characteristics or availability. Constructing a dataset set can therefore require both restrictions on eligible individual datasets and an objective governing the composition of the complete set.

The research gap therefore lies in integrating these existing capabilities. Existing work provides methods for dataset discovery, performance-based representation and exploration, and set-level optimization, but the reviewed approaches do not provide an integrated, recommendation-centered workflow that jointly supports required datasets, dataset-level constraints, configurable target-set sizes, alternative set-level objectives, result inspection, and documentation. FINALLY addresses this integration gap by combining established performance-based selection principles with these configuration, inspection, and export capabilities. Chapter~\ref{chap:methodology} describes the requirement derivation and system development.

% ============================================================================
% CHAPTER 3: METHODOLOGY
%
% Describe:
%
% - the research design,
% - the data,
% - the algorithms,
% - the implementation,
% - the experimental protocol, and
% - all decisions required to reproduce the work.
%
% Distinguish methodological decisions from implementation details.
% ============================================================================

\cleardoublepage

\chapter{Methodology}
\label{chap:methodology}

This chapter describes the methodological process used to develop and evaluate FINALLY. Section~\ref{sec:research-design} presents the overall research design and relates its main components to the research questions. Section~\ref{sec:iterative-development} then explains how the initial system requirements were derived, refined through iterative development, and verified before the final systematic evaluation. Chapter~\ref{chap:finally} presents the resulting system, while Chapter~\ref{chap:evaluation} describes the evaluation setup in detail.

\section{Research Design}
\label{sec:research-design}

This thesis uses a project-specific, artifact-centered, and iterative research design. The methodological process combines the derivation and refinement of system requirements, the implementation of a software artifact, and a subsequent technical and algorithmic evaluation. FINALLY constitutes the resulting artifact and operationalizes the proposed recommendation-centered workflow for constructing configurable dataset sets.

An artifact-centered approach was appropriate because the first research question concerns the realization of an integrated workflow rather than the comparison of existing systems. Developing a functional artifact made it possible to translate the identified requirements into an executable process and to examine how they interact within a complete system. The subsequent technical and algorithmic evaluation complemented this constructive component by testing whether the artifact produced dataset sets in accordance with its configured constraints and selection objectives.

To address the first research question, I derived the system requirements, developed FINALLY through iterative refinement, and examined the resulting workflow and system capabilities. The implementation itself does not constitute the complete answer. Rather, it provides the concrete artifact through which the proposed form of support can be demonstrated and analyzed.

To address the second research question, I systematically evaluated whether the generated dataset sets satisfy user-defined constraints and reflect the selected recommendation strategy across different configurations. The evaluation examines constraint compliance, strategy alignment, deterministic reproducibility, and end-to-end response time. Chapter~\ref{chap:evaluation} specifies the evaluated configurations, data basis, measurements, and analysis procedures.

The development did not follow a predefined software-development or design-science framework. Instead, it proceeded through iterative implementation and review cycles. Supervisory discussions, observations during implementation, and the capabilities of the existing APS Explorer infrastructure influenced the requirements and priorities of these iterations. Section~\ref{sec:iterative-development} describes how requirement derivation and system development influenced one another throughout this process.

This thesis evaluates observable technical and algorithmic properties of the implemented artifact. It does not include a user study and therefore makes no empirical claims about perceived usability, user satisfaction, or the effect of FINALLY on researchers' final dataset-selection decisions. Instead, the evaluation determines whether the system behaves according to its configured constraints and selection objectives under systematically varied conditions. Accordingly, the evaluation verifies FINALLY's implemented behavior but does not assess its practical use by researchers.

\section{Requirement Derivation and Iterative System Development}
\label{sec:iterative-development}

Requirement derivation and system development were interdependent rather than strictly sequential. An initial specification established the central recommendation workflow, while further requirements and implementation priorities emerged during the development and review of intermediate system versions.

A GitHub issue in the APS Explorer repository documented the initial workflow.\footnote{\url{https://github.com/ISG-Siegen/APS-Explorer-Website/issues/7}} It proposed that researchers should be able to retain previously selected datasets, define a desired total set size, restrict the available datasets through filters, receive recommendations for the remaining positions, inspect the resulting set in an \gls{aps}, and export the result. More complex extensions involving dataset popularity were explicitly assigned to a later version and were not included in the scope of this thesis.

The requirements were not treated as a fixed specification established before implementation. The initial issue defined the core workflow, whereas later requirements emerged when intermediate versions exposed additional needs for validation, documentation, inspection, and operational analysis. Requirements were therefore refined when implementation findings and supervisory feedback exposed additional needs, while the central dataset-set recommendation workflow remained unchanged.

The iterative process resulted in the following core requirements:

\begin{enumerate}
    \item FINALLY should allow researchers to define datasets that must remain part of the final dataset set.
    \item Researchers should be able to restrict the pool of eligible candidate datasets through explicit selection and metadata filters.
    \item The requested total dataset-set size should be configurable and validated against the available candidate pool.
    \item The system should support Random, diverse, and non-diverse dataset-set selection through configurable performance metrics and cutoffs.
    \item Generated dataset sets should be inspectable through dataset metadata and an \gls{aps}-based visualization.
    \item Recommendation configurations and results should be exportable for documentation and reuse.
    \item Submitted configurations and results should be recordable for administrative analysis.
\end{enumerate}

APS Explorer provided the technical starting point for my implementation. Its existing dataset and performance data allowed me to focus the initial prototype on the recommendation workflow rather than on constructing a new data infrastructure \cite{vente2025apsexplorer}. Building on this infrastructure, I implemented an initial \emph{Dataset Recommendation} tab within APS Explorer that supported \emph{Random} selection.

I subsequently extended the recommender through a series of project-specific iterations. The first extension added metadata filters and configuration validation. Following supervisory discussions, I implemented export functionality as the next development step. In a later iteration, I adapted and integrated the \gls{effcov} and Convex Hull objectives described by Reising \cite{reising2026systematic}. These existing objectives provided the basis for the diverse and non-diverse \gls{aps}-based strategies implemented in FINALLY.

In subsequent iterations, I implemented usage logging, redesigned the public interface, and developed a protected administrative interface. As the recommendation workflow evolved into an application with a distinct purpose and interaction model, I separated it from the original APS Explorer frontend and named the resulting system FINALLY. During the final development stages, I added result inspection and further presentation refinements. Chapter~\ref{chap:finally} describes the resulting functions and architecture in detail.

Recurring supervisory discussions provided checkpoints for reviewing intermediate versions and determining the priorities of subsequent iterations. The intermediate versions were also examined through complete executions of the recommendation workflow. Findings from these reviews led to corrections and smaller refinements before the evaluation phase, while additional numerical issues identified during an initial evaluation execution were corrected before the final systematic evaluation.

Development-time verification and the final systematic evaluation served different purposes. I used manual end-to-end tests for most user-facing functions and automated tests for the dataset-set selection procedures and the generation of export artifacts. These checks were used to detect functional defects and confirm that individual components behaved as intended. Findings from an initial evaluation execution were likewise treated as implementation feedback when they exposed numerical defects in the Convex Hull selection procedure. These verification activities did not directly answer the research questions. The final systematic evaluation instead applied predefined configurations and measurements to the corrected and fixed implementation, as described in Chapter~\ref{chap:evaluation}.

An initial execution of the systematic evaluation revealed numerical defects in the Convex Hull selection procedure. The intrinsic-rank estimation could retain a numerical component beyond the theoretical maximum rank of \(n-1\) for \(n\) centered vectors, the hull-volume calculation could return a zero volume for valid high-dimensional minimal simplices, and the score-comparison logic used an absolute tolerance floor that could treat distinct very small Convex Hull scores as ties. These findings were treated as implementation defects rather than as final evaluation results.

I corrected the identified defects before conducting the final evaluation. I revised the intrinsic-rank estimation to enforce the theoretical maximum rank of \(n-1\), introduced a determinant-based volume calculation for high-dimensional minimal simplices, and replaced the absolute tolerance floor used for score comparisons with a relative tolerance at the magnitude of the compared values. I then used targeted regression tests and targeted verification to confirm the corrected behavior before repeating the complete evaluation. Section~\ref{sec:finally-selection} describes the resulting selection implementation in detail.

Commit~\texttt{48943683}\footnote{Full commit hash: \nolinkurl{489436834f95300c84bcaf3843b3c472eb1e8be3}.} defines the FINALLY implementation used for the final systematic evaluation. After this implementation was fixed, the complete 420-run experiment was repeated using the predefined evaluation design. Only the results of this corrected final evaluation are used to answer the research questions and are reported in Chapters~\ref{chap:results}--\ref{chap:conclusion}. The initial evaluation is retained separately as a provenance artifact. This commit therefore identifies the exact FINALLY source state to which the reported evaluation results refer.

% State relevant software versions, libraries, hardware, random seeds, and
% implementation decisions. Refer to source code in the appendix or an
% external repository where appropriate.

% ============================================================================
% CHAPTER 4: FINALLY
% ============================================================================

\chapter{FINALLY}
\label{chap:finally}

This chapter presents the implemented FINALLY system. Section~\ref{sec:finally-overview} introduces its purpose, scope, and principal use cases. Section~\ref{sec:finally-architecture} describes the system architecture and the responsibilities of its main components. Section~\ref{sec:finally-workflow} then explains the user-facing recommendation workflow, while Section~\ref{sec:finally-selection} details how FINALLY constructs dataset sets according to the selected recommendation strategy. Finally, Section~\ref{sec:finally-results} describes how generated dataset sets can be inspected and exported. Chapter~\ref{chap:methodology} describes the development process and the derivation of the system requirements. Chapter~\ref{chap:evaluation} presents the evaluation design.

\section{System Overview}
\label{sec:finally-overview}

To address the research objective, I developed FINALLY, an open-source, web-based dataset recommender designed to support researchers in constructing dataset sets for offline recommender-systems evaluations. At the time of writing, the system provides access to 96 recommender-systems datasets and is publicly available as a deployed web application\footnote{\url{https://finally.recommender-systems.com}}. Its source code is published separately\footnote{\url{https://code.isg.beel.org/FINALLY}}. FINALLY builds on the code base\footnote{\url{https://code.isg.beel.org/APS-Explorer-Website}} and data infrastructure of APS Explorer \cite{vente2025apsexplorer}, but reorganizes the interaction around the generation of configurable dataset sets rather than the manual exploration of individual datasets.

Users configure a recommendation by specifying an optional set of existing datasets, a desired final set size, restrictions on the candidate pool, and a recommendation strategy. For \gls{aps}-based strategies, users additionally select the performance metric and cutoff from which the relevant performance representation is constructed. The interface refers to datasets that must remain part of the generated set as \emph{Seed Datasets}. Functionally, these datasets constitute required components of the final result. To avoid confusion with random-number-generator seeds, they are therefore referred to as \emph{required datasets} throughout this thesis, except when a concrete interface element is described.

Based on this configuration, FINALLY generates a complete dataset set containing the required datasets and the newly recommended datasets. The result can be inspected through dataset metadata and an \gls{aps}-based visualization and can subsequently be exported for experimental documentation and publications.

FINALLY supports three principal use cases. First, a researcher who has not yet selected any datasets can generate an initial dataset set according to a chosen recommendation strategy. This use case supports the early planning of an evaluation before implementation, experimentation, and algorithm tuning begin.

Second, a researcher can extend an existing dataset selection. Previously selected datasets are entered as required datasets, while FINALLY recommends the remaining datasets needed to reach the requested target-set size. This allows an evaluation to be extended either with datasets that differ in their algorithm-performance patterns, with datasets that exhibit comparatively similar patterns, or with randomly selected datasets, depending on the selected strategy.

Third, a researcher who has already developed, tested, and tuned an algorithm can request additional datasets that were not used during development. The generated dataset set can then support a post-development validation in which the algorithm is examined under additional empirical conditions. FINALLY cannot verify whether these datasets were actually unseen during development; this depends on the researcher's preceding experimental process.

In addition to the public recommendation interface, I implemented an internal administrative interface for inspecting recorded usage data, submitted recommendation configurations, and the corresponding results.

FINALLY is a decision-support system for dataset selection rather than a complete recommender-systems experimentation platform. It neither trains nor evaluates recommendation algorithms on the generated dataset sets. Furthermore, compliance with the configured target-set size, filters, and recommendation objective does not guarantee that every selected dataset is suitable for a particular experimental protocol. Researchers remain responsible for assessing aspects such as domain relevance, preprocessing compatibility, licensing conditions, evaluation methodology, and the relation between the selected datasets and the research question.

\section{System Architecture}
\label{sec:finally-architecture}

\begin{figure}[htbp]
    \centering
    \includegraphics[width=\textwidth]{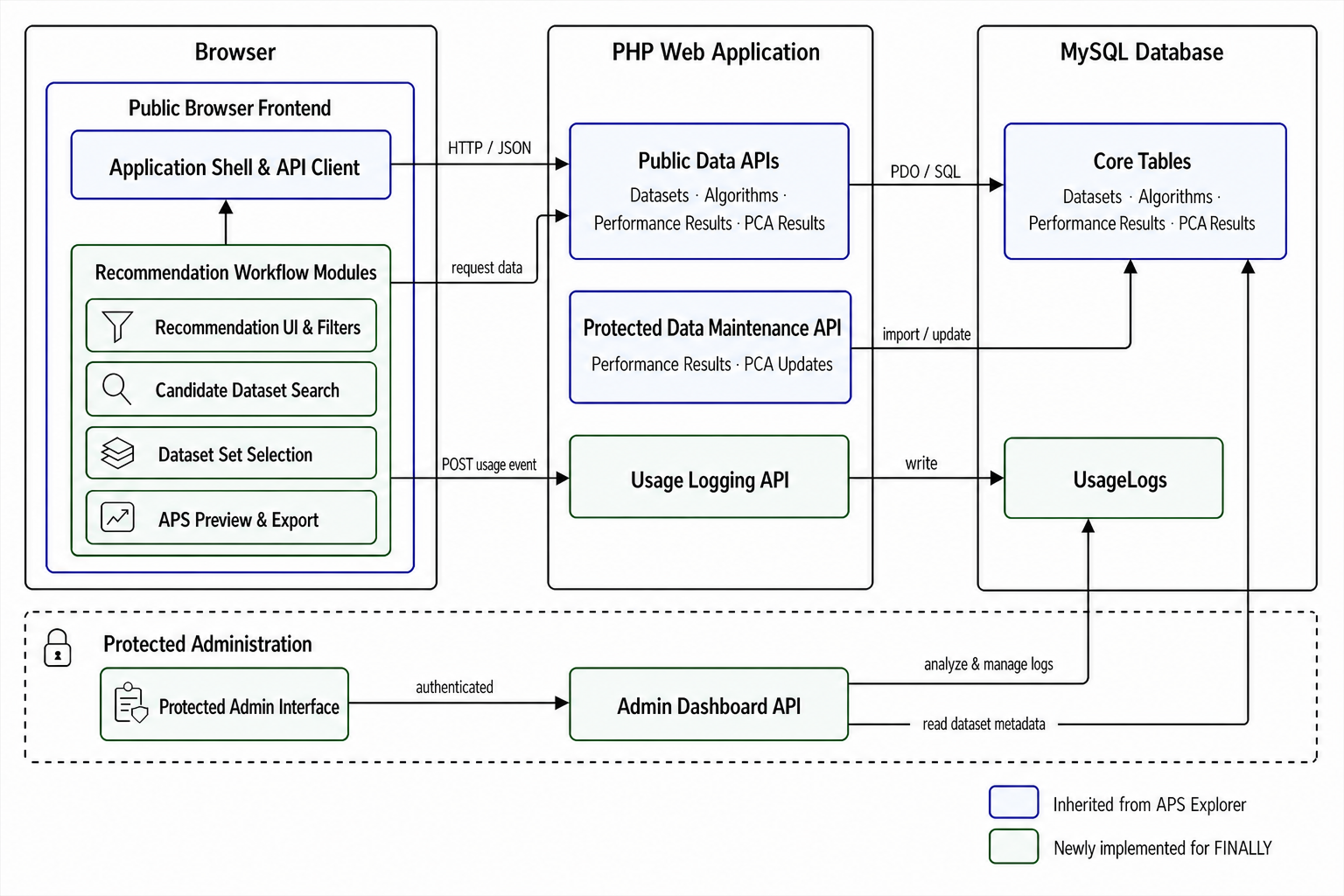}
    \caption{High-level architecture of FINALLY. Blue components were inherited from APS Explorer, whereas green components were newly implemented for FINALLY.}
    \label{fig:finally-architecture}
\end{figure}

FINALLY follows a browser-centered web architecture consisting of a public frontend, PHP-based \glspl{api}, and a MySQL database. The architecture separates the public recommendation workflow from server-side data access, usage logging, and the protected administrative interface. Figure~\ref{fig:finally-architecture} provides a high-level overview of these components and distinguishes the infrastructure inherited from APS Explorer from the components newly implemented for FINALLY.

The public frontend contains the complete user-facing recommendation workflow. After loading the required data through the inherited \glspl{api}, it validates the user configuration, constructs the candidate pool, and executes the selected dataset-set selection strategy in the browser. The filtering logic, recommendation procedures, result table, visualization, sharing functionality, and exports are therefore client-side components. No separate server-side recommendation service is used.

Completed recommendations can additionally be stored through a newly implemented usage-logging \gls{api}. Each record contains the submitted configuration and the generated final dataset set, including the selected \emph{Seed Datasets}, candidate pool, applied filters, requested target-set size, recommendation strategy, metric, and cutoff. These records provide the data basis for the administrative interface.

The administrative interface is separate from the public recommendation workflow and serves as an auxiliary component for operational analysis. It accesses stored usage records through dedicated PHP endpoints and presents aggregated information and visualizations derived from them. The complete administrative directory, including its \gls{api} endpoints, is protected through an \texttt{.htaccess} configuration. The interface supports the inspection of recorded configurations, generated recommendations, and aggregate usage information, but does not provide functionality for importing performance data or recalculating dataset representations. This separation keeps administrative analysis distinct from the user-facing dataset-selection process.

The reuse of APS Explorer is concentrated in the data layer. In developing FINALLY, I retained the existing MySQL database and public PHP \glspl{api}, while replacing most of the original frontend and its page-navigation structure. I implemented the recommendation interface, filtering and selection logic, sharing functionality, result presentation, exports, usage logging, and administrative interface specifically for FINALLY. I also reimplemented the \gls{aps} visualization, while retaining its underlying interaction concept from APS Explorer.

\section{Recommendation Workflow}
\label{sec:finally-workflow}

\begin{figure}[htbp]
    \centering
    \includegraphics[
        width=\textwidth,
        height=0.72\textheight,
        keepaspectratio
    ]{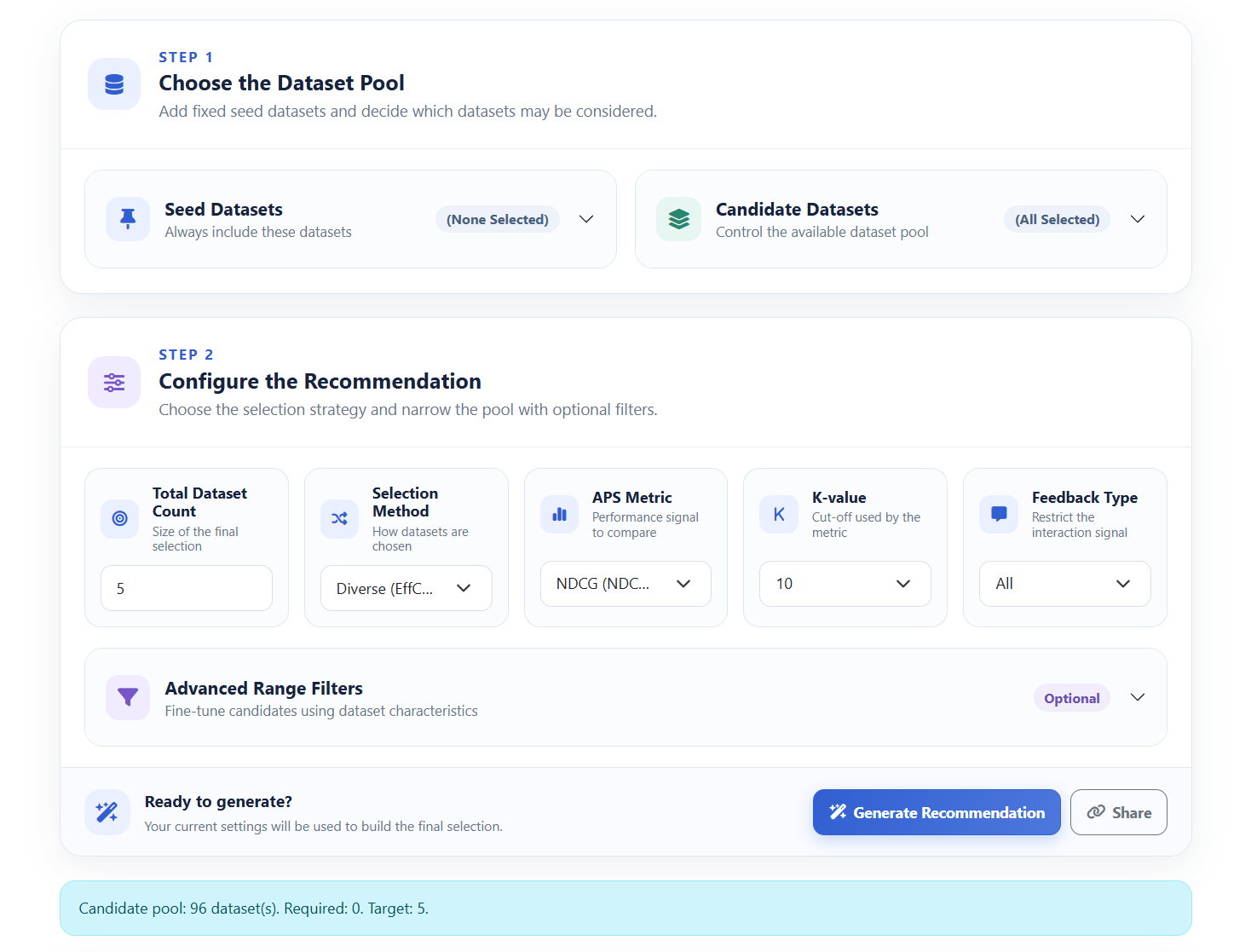}
    \caption{Configuration stage of FINALLY. Step~1 defines the required and candidate datasets, while Step~2 specifies the target-set size, recommendation strategy, performance configuration, and optional metadata filters. The status area summarizes and validates the current request.}
    \label{fig:finally-configuration}
\end{figure}

I designed FINALLY's user-facing recommendation process as three consecutive steps on a single page: selecting the dataset pool, configuring the recommendation, and inspecting the generated dataset set. This structure keeps the initial dataset selection, the recommendation parameters, and the resulting set within one continuous workflow. Figure~\ref{fig:finally-configuration} shows the first two steps and the status information displayed before a recommendation is generated.

In Step~1, \emph{Choose the Dataset Pool}, users define the starting point and the eligible candidate datasets. The \emph{Seed Datasets} field contains datasets that must be retained in the final set. As explained in Section~\ref{sec:finally-overview}, these are referred to as required datasets throughout this thesis. Selecting \emph{Seed Datasets} is optional, which allows FINALLY to either generate an entirely new dataset set or extend an existing selection.

The adjacent \emph{Candidate Datasets} field determines which datasets may be newly recommended. All 96 available datasets are selected by default, but users can explicitly include or exclude individual datasets. The resulting candidate pool can subsequently be restricted further through the filters provided in Step~2. Required datasets and candidate datasets therefore serve different purposes: the former remain fixed components of the final set, whereas the latter define the datasets from which additional recommendations may be drawn.

Step~2, \emph{Configure the Recommendation}, specifies the properties of the requested dataset set and the recommendation strategy. The \emph{Total Dataset Count} defines the size of the complete result, including the selected \emph{Seed Datasets}. The interface enforces a value that is at least one greater than the number of \emph{Seed Datasets}, ensuring that each execution adds at least one newly recommended dataset.

Users select one of five recommendation methods: \emph{Diverse (\gls{effcov})}, \emph{Diverse (ConvexHull)}, \emph{Non-Diverse (\gls{effcov})}, \emph{Non-Diverse (ConvexHull)}, or \emph{Random}. The diverse and non-diverse alternatives determine whether the corresponding \gls{aps}-based score is maximized or minimized, while \emph{Random} provides an undirected selection. The internal construction procedures underlying these methods are described in Section~\ref{sec:finally-selection}.

The \emph{\gls{aps} Metric} and \emph{K-value} controls select the performance signal used by the \gls{aps}-based strategies and the corresponding stored \gls{aps} visualization. FINALLY supports \gls{ndcg}, \gls{hr}, and Recall at cutoffs 1, 3, 5, 10, and 20. These controls remain visible when \emph{Random} is selected. In this case, they do not determine the random ordering of the candidate datasets, but they continue to determine the \gls{aps} representation used to inspect the generated set.

Candidate datasets can additionally be restricted by feedback type and dataset characteristics. At the time of writing, FINALLY contains only datasets with implicit feedback. The \emph{Feedback Type} control therefore does not currently distinguish between multiple available feedback types. The optional \emph{Advanced Range Filters} allow users to specify minimum and maximum values for properties including the number of interactions, users, and items, the user--item ratio, density, and several user- and item-level interaction statistics. The controls combine range sliders with numerical inputs, allowing either approximate visual adjustment or direct entry of threshold values.

The filters apply to the candidate pool from which new recommendations are selected. They do not remove the \emph{Seed Datasets} that the user has explicitly required as part of the final set. Thus, the filters constrain only newly recommended datasets and do not remove required datasets from the final set.

A status area between the configuration and result sections provides immediate feedback about the current request. Under a valid configuration, it summarizes the size of the candidate pool, the number of required datasets, and the requested target-set size. If the configured filters and candidate selection leave too few eligible datasets, the same area displays a validation message, such as \emph{Not enough candidate datasets to reach your target count with current filters}. Invalid configurations are therefore identified before an incomplete dataset set is presented.

The user starts the process through the \emph{Generate Recommendation} button. During execution, the status area displays a loading indicator and informs the user that the recommendation is being generated. Once the procedure has completed, the resulting dataset set appears directly in Step~3, \emph{Final Dataset Selection}, on the same page. The result table, \gls{aps} visualization, and export functionality are described in Section~\ref{sec:finally-results}.

Additional interface states, including the expanded \emph{Advanced Range Filters}, are shown in Appendix~\ref{app:interface-views}.

\section{Dataset-Set Selection}
\label{sec:finally-selection}

FINALLY transforms the validated configuration described in Section~\ref{sec:finally-workflow} into a complete dataset set. The selection procedure receives the required datasets, the filtered pool of eligible candidates, the requested total dataset count, and the selected recommendation method. Required datasets remain fixed, while FINALLY selects the number of additional candidates needed to reach the requested total size.

The five available recommendation methods are summarized in Table~\ref{tab:finally-strategies}. The theoretical motivation and mathematical definitions of \gls{aps}-based diversity are presented in Section~\ref{sec:related-work-diversity}. The following description focuses on their operational implementation in FINALLY.

\begin{table}[htbp]
    \centering
    \small
    \caption{Dataset-set selection methods implemented in FINALLY.}
    \label{tab:finally-strategies}
    \begin{tabular}{
        p{0.35\textwidth}
        p{0.22\textwidth}
        p{0.14\textwidth}
        p{0.15\textwidth}
    }
        \toprule
        Interface option & Selection objective & Direction & Deterministic \\
        \midrule

        Diverse (\gls{effcov})
        & \gls{effcov} score
        & Maximize
        & Yes \\

        Diverse (ConvexHull)
        & Convex Hull score
        & Maximize
        & Yes \\

        Non-Diverse (\gls{effcov})
        & \gls{effcov} score
        & Minimize
        & Yes \\

        Non-Diverse (ConvexHull)
        & Convex Hull score
        & Minimize
        & Yes \\

        Random
        & Random sampling
        & --
        & No \\

        \bottomrule
    \end{tabular}
\end{table}

\paragraph{Performance-vector construction.}

The \gls{aps}-based strategies do not operate on the two-dimensional coordinates displayed in the result visualization. Instead, FINALLY constructs a high-dimensional performance vector for each eligible dataset. Each dimension corresponds to one available recommendation algorithm, ordered consistently by algorithm name, and contains its performance for the metric and cutoff selected in the interface. Consequently, the selection methods compare datasets according to their complete algorithm-performance patterns rather than their positions in a two-dimensional projection.

When multiple performance configurations are available for the same combination of dataset, algorithm, metric, and cutoff, FINALLY uses their arithmetic mean. A dataset participates in \gls{aps}-based selection if at least one algorithm has a finite value for the selected metric and cutoff. At the time of writing, all 96 datasets satisfy this condition for all 15 supported metric--cutoff combinations.

FINALLY imputes missing dimensions with the mean of the corresponding algorithm column across the imputation domain. This domain comprises the eligible candidate pool and the required datasets with usable performance data. If an algorithm column contains no finite value within this domain, FINALLY replaces the complete column with zero. FINALLY retains the raw performance values and applies no feature-wise min--max normalization or standardization before dataset-set selection. All dimensions within a given selection problem therefore remain on the original scale of the same performance metric and cutoff. This design preserves the observed differences in algorithm performance without introducing an additional feature-scaling transformation, but it also means that algorithm dimensions with greater variation can contribute more strongly to Euclidean distances, covariance, and the resulting selection scores.

\paragraph{Implemented objective functions.}

The Convex Hull and \gls{effcov} objectives I implemented in FINALLY are adapted from the \gls{aps}-based dataset-set diversity measures introduced by Reising \cite{reising2026systematic}. For a selected dataset set \(S\), FINALLY calculates the raw objectives
\begin{equation}
    D_{\mathrm{FINALLY}}^{\mathrm{CH}}(S)
    =
    \operatorname{MNND}(S)
    \cdot
    \operatorname{CH}_{\mathrm{intr}}(S)
\end{equation}
and
\begin{equation}
    D_{\mathrm{FINALLY}}^{\mathrm{EffCov}}(S)
    =
    \operatorname{MNND}(S)
    \cdot
    \operatorname{EffCov}_{\mathrm{struct}}(S).
\end{equation}
For both objectives, the diverse strategy maximizes the corresponding score, whereas the non-diverse strategy minimizes it. No separate similarity measure is used.

For \gls{effcov}, a set containing fewer than two performance vectors receives a score of zero. Otherwise, FINALLY centers the \(n\) selected performance vectors as defined in Section~\ref{sec:related-work-diversity} and computes covariance eigenvalues equivalent to those of the sample covariance matrix
\begin{equation}
    C
    =
    \frac{1}{n-1}X^{\mathsf{T}}X,
\end{equation}
where \(X\) denotes the centered performance matrix. Thus, FINALLY uses the sample-covariance convention with denominator \(n-1\). The implementation obtains these eigenvalues from the singular values \(\sigma_i\) of \(X\) as
\begin{equation}
    \lambda_i
    =
    \max\left(
        0,
        \frac{\sigma_i^2}{n-1}
    \right),
\end{equation}
and retains an eigenvalue if
\begin{equation}
    \lambda_i
    >
    10^{-12}\lambda_{\max},
\end{equation}
where \(\lambda_{\max}\) is the largest covariance eigenvalue. The number of retained eigenvalues defines \(d_{\mathrm{eff}}\). If \(\lambda_{\max}\) is non-positive or non-finite, or if \(d_{\mathrm{eff}}=0\), FINALLY returns an \gls{effcov} score of zero. Otherwise, the structure term is evaluated in logarithmic form as
\begin{equation}
    \operatorname{EffCov}_{\mathrm{struct}}(S)
    =
    \exp\left(
        \frac{1}{2d_{\mathrm{eff}}}
        \sum_{i=1}^{d_{\mathrm{eff}}}
        \ln \lambda_i
    \right).
\end{equation}
This is mathematically equivalent to the geometric mean of the retained standard deviations while avoiding the direct multiplication of potentially very small eigenvalues. If the resulting structure term is non-finite, FINALLY treats it as zero. The final score is the product of this structure term and \(\operatorname{MNND}(S)\).

For the Convex Hull objective, FINALLY first centers the selected performance vectors and projects them onto their numerically estimated intrinsic affine subspace. The centered matrix is centered again before the rank estimation to remove residual offsets introduced by floating-point arithmetic. Let \(X \in \mathbb{R}^{n \times d}\) denote this re-centered matrix. FINALLY obtains its singular values from an eigendecomposition of the smaller Gram matrix \(XX^{\mathsf{T}}\) if \(n \leq d\), and \(X^{\mathsf{T}}X\) otherwise.

Let \(G\) denote this Gram matrix, \(g\) its dimension, \(l\) the length of
the dot products used to construct it, and
\(s_G = \max_{i,j} |G_{ij}|\). The Jacobi eigendecomposition uses the
tolerance \(t_J = 10^{-12}s_G\) for \(s_G > 0\), and
\(t_J = 10^{-12}\) for \(s_G = 0\). Let \(\lambda_{\max}\) denote the
largest estimated eigenvalue of \(G\), and define
\[
s_{\mathrm{num}}
=
\max\bigl(s_G,\max(0,\lambda_{\max})\bigr).
\]
For \(s_{\mathrm{num}} > 0\), FINALLY estimates the numerical eigenvalue
uncertainty as
\[
t_{\lambda}
=
\max\Bigl(
(g-1)t_J,\,
\epsilon_{\mathrm{mach}}
\max(1,g,l)s_{\mathrm{num}}
\Bigr),
\]
where \(\epsilon_{\mathrm{mach}}\) denotes JavaScript's machine epsilon.
For \(s_{\mathrm{num}} = 0\), FINALLY sets \(t_{\lambda}=0\).

A singular value is considered rank-forming if
\[
\sigma_i > \sqrt{t_{\lambda}}.
\]
Let \(r_{\mathrm{num}}\) denote the number of singular values satisfying this
condition. FINALLY then determines the intrinsic rank as
\[
r
=
\min\bigl(r_{\mathrm{num}}, n-1, d\bigr).
\]
This explicitly enforces the theoretical maximum rank of \(n-1\) for \(n\)
centered vectors.

The intrinsic hull volume is calculated according to the resulting rank. A rank of zero produces a hull volume of zero. For \(r=1\), FINALLY uses the length of the interval between the minimum and maximum projected coordinates, and for \(r=2\), it calculates the area of the two-dimensional convex hull. For \(r\geq3\), a set containing exactly \(r+1\) projected points forms a minimal \(r\)-dimensional simplex. FINALLY calculates its volume directly from the edge matrix \(E\) as
\begin{equation}
    \operatorname{Vol}_{r}(S)
    =
    \frac{|\det(E)|}{r!}.
\end{equation}
For higher-dimensional projected sets containing more than \(r+1\) points, FINALLY uses its facet-based convex-hull construction.

The objectives I implemented in FINALLY deliberately differ from the complete metric definitions presented in Section~\ref{sec:related-work-diversity} in two respects. First, FINALLY does not apply Reising's set-size-specific normalization \(D^*(k)\). Within a fixed selection problem, omitting this normalization does not change the ordering of feasible sets because every score would be divided by the same positive constant, but the resulting raw scores should not be compared directly across configurations with different performance spaces, candidate pools, or target-set sizes. Second, FINALLY omits the additive stabilizer \(\varepsilon\). This difference can affect the ordering of candidate sets when the corresponding structure term is zero because the \gls{mnnd} contribution is then also eliminated. The implemented objectives should therefore be understood as adaptations of Reising's measures rather than exact reproductions of their complete normalized definitions.

\paragraph{Numerical score comparison.}

The greedy search must compare objective values that can differ by many orders of magnitude, particularly for high-dimensional Convex Hull volumes. FINALLY therefore compares two finite selection scores \(s_1\) and \(s_2\) using the scale-dependent tolerance
\begin{equation}
    \tau(s_1,s_2)
    =
    10^{-12}
    \max\left(
        |s_1|,
        |s_2|
    \right).
\end{equation}
The scores are treated as numerically equal only if their absolute difference does not exceed this tolerance. Unlike an absolute tolerance with a fixed lower bound, this comparison remains relative to the magnitude of the scores and therefore allows distinct values substantially below \(10^{-12}\) to influence the selection. FINALLY applies the same comparison when choosing the next greedy candidate and when selecting the best final solution across multiple starting points. Numerically equal alternatives are resolved through deterministic index- and order-based tie-breaking rules.

\paragraph{Greedy dataset-set construction.}

FINALLY does not enumerate all possible candidate subsets. Instead, it applies a greedy multi-start heuristic. During each greedy run, the current set is extended one dataset at a time. At every step, FINALLY temporarily adds each remaining candidate to the current set and calculates the score of the resulting set. FINALLY retains the candidate that produces the best score in the selected optimization direction. This process continues until the requested number of additional datasets has been selected.

When no required datasets are present and more than one dataset must be recommended, FINALLY performs this greedy construction from multiple starting candidates. It generates up to 60 deterministic start points. An internal pseudo-random generator initialized with the fixed value \(0\) selects the first start. A farthest-first procedure then selects the remaining starts based on their distances from the previously selected starting points. Each start produces one complete greedy solution, after which FINALLY returns the solution with the best final score. If the candidate pool contains no more than 60 datasets, each candidate can serve as a starting point. For larger pools, the search is restricted to 60 starts.

When at least one required dataset is present, these datasets form the fixed initial set. FINALLY then performs one greedy run beginning from that set rather than executing the multi-start procedure. Each candidate is evaluated together with the required datasets and the candidates already added during the current run. Required datasets can therefore directly influence which additional datasets optimize the selected score.

A special case occurs when no required dataset is present and only one dataset must be recommended. The implemented objective functions assign a score of zero to singleton sets and therefore cannot distinguish between individual candidates. For a diverse strategy, FINALLY selects the candidate farthest from the centroid of the candidate pool. For a non-diverse strategy, it selects the candidate nearest to this centroid. When the requested number of new datasets is at least as large as the remaining candidate pool, all eligible candidates are returned without evaluating alternative subsets.

Because candidates are selected sequentially from a limited number of starting points, the procedure does not guarantee a globally optimal dataset set.

\paragraph{Required datasets and exceptional cases.}

For \gls{aps}-based selection, required datasets with usable performance data are included as fixed vectors in every candidate evaluation. If a required dataset has no finite performance value for any algorithm at the selected metric and cutoff, it remains part of the final result but does not participate in the Convex Hull or \gls{effcov} optimization. Required datasets with partially missing performance vectors remain usable through the imputation procedure described above.

FINALLY prevents duplicate datasets in the final set, and a required dataset cannot be recommended again. If the eligible candidate pool is insufficient, FINALLY does not execute the selection and instead applies the validation behavior described in Section~\ref{sec:finally-workflow}. FINALLY does not replace an unavailable \gls{aps}-based recommendation with \emph{Random} selection.

\paragraph{Randomness and determinism.}

The four \gls{aps}-based methods are deterministic for an unchanged ordered input. Their multi-start initialization uses the fixed internal seed \(0\), candidate iteration follows a stable order, and numerically equal objective scores are resolved through deterministic ordering rules. The scale-aware score comparison described above is applied consistently during greedy candidate selection and when comparing the completed multi-start solutions. Repeated executions with the same data, input ordering, and configuration are therefore expected to produce the same dataset membership and order, subject to the deterministic floating-point behavior of the numerical calculations.

Random selection shuffles the eligible candidate list using JavaScript's built-in pseudorandom number generator \texttt{Math.random()}\footnote{\url{https://developer.mozilla.org/en-US/docs/Web/JavaScript/Reference/Global_Objects/Math/random}} and returns the required number of candidates from the resulting order. FINALLY does not expose a configurable random-number-generator seed. Repeated \emph{Random} executions can therefore return different dataset sets. Required datasets remain fixed but do not influence the random ordering of the eligible candidates.

\section{Result Inspection and Export}
\label{sec:finally-results}

\begin{figure}[p]
    \centering

    \begin{subfigure}{\textwidth}
        \centering
        \includegraphics[
            width=\textwidth,
            height=0.42\textheight,
            keepaspectratio
        ]{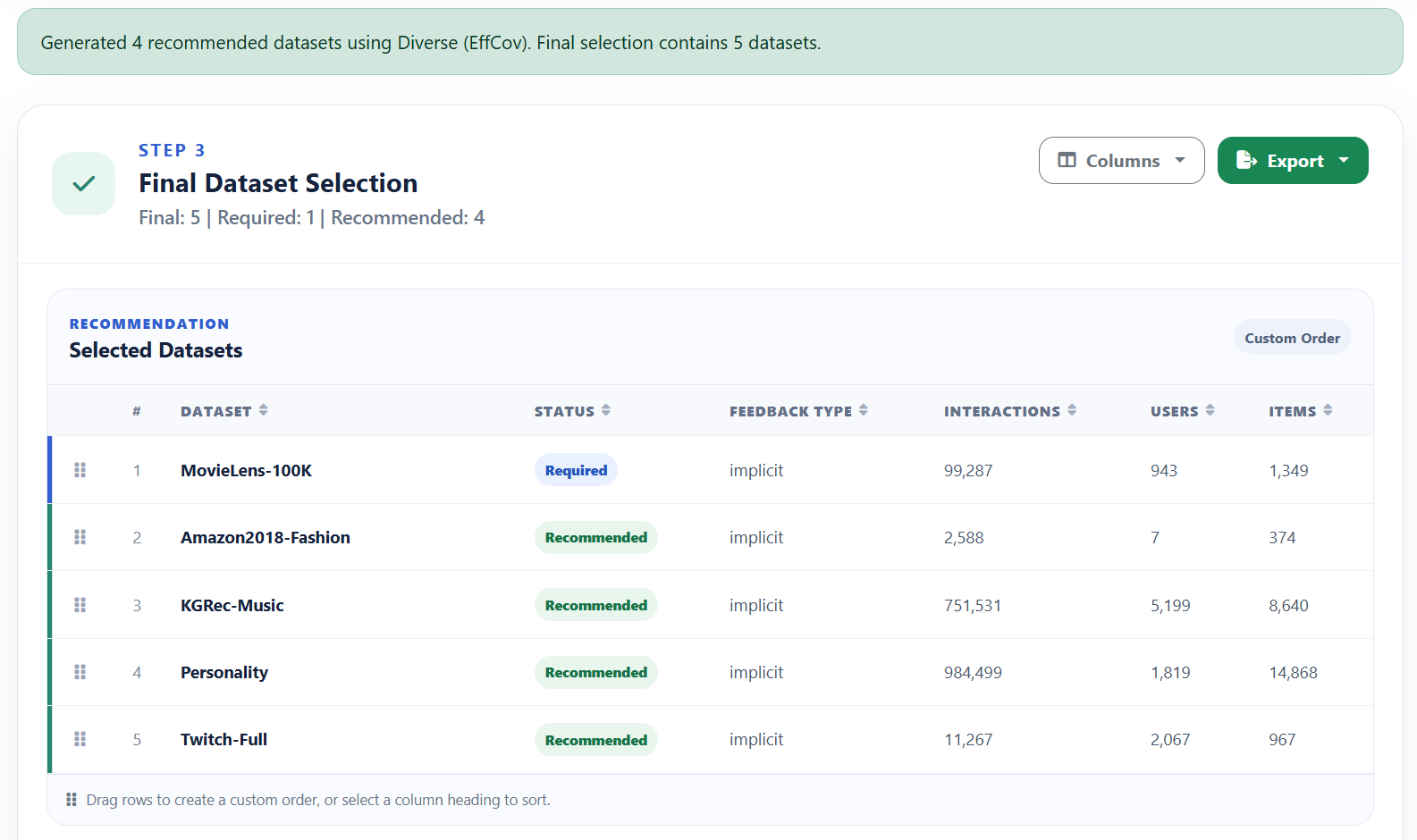}
        \caption{Result table in Step~3 of FINALLY. The table distinguishes
        required datasets from newly recommended datasets and provides controls
        for configuring the visible columns and exporting the generated dataset set.}
        \label{fig:finally-result-table}
    \end{subfigure}

    \vspace{0.8em}

    \begin{subfigure}{\textwidth}
        \centering
        \includegraphics[
            width=0.75\textwidth,
            height=0.28\textheight,
            keepaspectratio
        ]{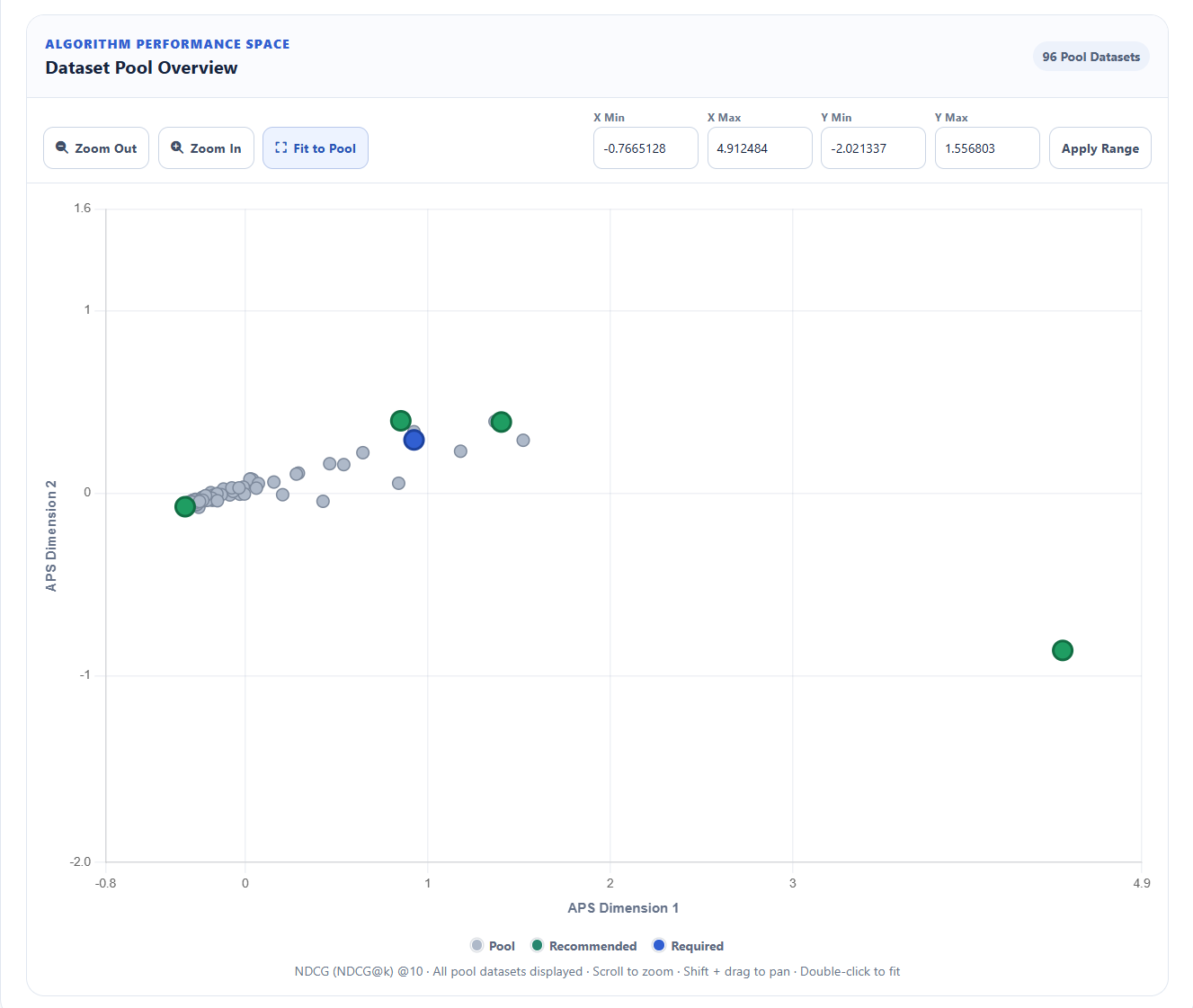}
        \caption{Two-dimensional \gls{aps} visualization of a generated dataset
        set. Candidate-pool datasets are shown in gray, recommended datasets in
        green, and required datasets in blue.}
        \label{fig:finally-aps-visualization}
    \end{subfigure}

    \caption{Result inspection in FINALLY.}
    \label{fig:finally-result-inspection}
\end{figure}

After a recommendation has been generated, FINALLY presents the completed dataset set in Step~3, \emph{Final Dataset Selection}. The result view combines a configurable dataset table with a two-dimensional representation of the corresponding \gls{aps}. This allows researchers to inspect both the descriptive properties of the selected datasets and their relative positions within the displayed performance space before exporting the result. Figure~\ref{fig:finally-result-table} shows the result table and its associated controls, while Figure~\ref{fig:finally-aps-visualization} shows the corresponding \gls{aps} visualization.

The result table distinguishes between datasets selected through the \emph{Seed Datasets} field and datasets added by the recommendation procedure. The former are labeled \emph{Required}, whereas the latter are labeled \emph{Recommended}. The row number, dataset name, and status remain permanently visible.

Through the \emph{Columns} control, users can display or hide the dataset properties available in the filtering interface. This allows the result table to focus on the metadata relevant to the current experimental setting. The status labels make it possible to distinguish the fixed components of the configuration from the datasets added by FINALLY. At the same time, the configurable metadata columns allow researchers to compare the properties of the selected datasets directly within the generated result.

All table columns can be used for sorting. Users can also define a custom dataset order by moving rows through drag-and-drop or the corresponding keyboard interaction. The resulting order is retained by the export functions, allowing the displayed dataset sequence to be transferred into subsequent documentation. Dataset membership itself cannot be changed within the result view. Adding or removing datasets requires adjusting the recommendation configuration and generating a new result.

The table is linked interactively to the \gls{aps} visualization. Hovering over a table row highlights the corresponding dataset point. Hovering over or clicking an \gls{aps} point highlights the corresponding table row, while hovering also displays a tooltip containing the dataset name, status, and two-dimensional coordinates.

The visualization contains the configured candidate pool and the datasets in the final result. Pool datasets are displayed in gray, datasets labeled \emph{Recommended} in green, and datasets labeled \emph{Required} in blue. Datasets that belong to neither the final result nor the candidate pool are not displayed. The visualization uses the metric and cutoff selected in the configuration interface and updates when the user changes this setting.

As explained in Section~\ref{sec:finally-selection}, the displayed two-dimensional coordinates support inspection but do not determine the internal optimization of the Convex Hull and \gls{effcov} strategies. Distances in the projected space represent differences in algorithm-performance patterns rather than general dataset similarity \cite{beel2024aps,vente2025apsexplorer}. Accordingly, proximity alone should not be interpreted as evidence that datasets are interchangeable. Researchers should interpret the visualization together with the dataset metadata and the intended experimental setting.

I implemented five export formats for FINALLY, summarized in Table~\ref{tab:finally-exports}. The \gls{png}, Markdown, \gls{html}, and LaTeX exports reflect the generated dataset set, its custom row order, and the metadata columns visible in the result table. The BibTeX export instead records the configuration and ordered final selection as provenance metadata.

The \gls{png} export preserves the report as a single visual artifact. Markdown, \gls{html}, and LaTeX provide structured textual representations. The \gls{html} format creates a standalone document with an embedded \gls{aps} visualization, while the LaTeX format creates a complete compilable document. The BibTeX export generates one provenance record for the complete FINALLY recommendation rather than bibliographic entries for individual datasets. It records the generation date, FINALLY \gls{url}, configuration, dataset counts, filters, and ordered final selection.

The export dialog previews the generated artifact. It displays text-based exports as source text and presents the \gls{png} export visually. Users can copy or download the generated content. The preview, clipboard content, and downloaded artifact use the same generated representation and therefore remain consistent.

FINALLY additionally provides a \emph{Share} function for transferring a recommendation configuration through a \gls{url}. The \gls{url} preserves the selected \emph{Seed Datasets}, \emph{Candidate Datasets}, filter settings, target-set size, strategy, metric, and cutoff, but not the generated result or a manually defined result order. Opening the \gls{url} therefore restores the inputs from which a new recommendation can be generated. A deterministic strategy should reproduce the same dataset membership if the underlying data, input ordering, and implementation remain unchanged. \emph{Random} may return a different dataset set because the \gls{url} does not preserve the original random outcome.

\begin{table}[htbp]
    \centering
    \small
    \caption{Export formats provided by FINALLY.}
    \label{tab:finally-exports}
    \begin{tabular}{
        p{0.16\textwidth}
        p{0.48\textwidth}
        p{0.24\textwidth}
    }
        \toprule
        Format & Exported content & Intended use \\
        \midrule

        \gls{png}
        & Styled report containing the configuration, filters, visible result table, and \gls{aps} visualization
        & Visual documentation \\

        Markdown
        & Textual report containing the summary, configuration, filters, result table, and \gls{aps} coordinate table
        & Repositories and Markdown-based reports \\

        \gls{html}
        & Standalone styled document containing the report, embedded \gls{aps} visualization, and coordinate table
        & Portable web-based documentation \\

        LaTeX
        & Standalone article source containing the summary, configuration, filters, result table, and \gls{aps} coordinate table
        & Academic reports and publications \\

        BibTeX
        & Single provenance entry describing the recommendation configuration and ordered final dataset set
        & Provenance and archival metadata \\

        \bottomrule
    \end{tabular}
\end{table}

Additional interface states, including the Markdown export dialog, are shown in Appendix~\ref{app:interface-views}.

% CHAPTER 5: EVALUATION
%
% Define:
%
% - datasets,
% - baselines,
% - evaluation metrics,
% - hypotheses,
% - experimental conditions,
% - statistical procedures, and
% - reproducibility measures.
%
% Explain why each metric and baseline is appropriate for the research
% question.
% 

\cleardoublepage

\chapter{Evaluation}
\label{chap:evaluation}

This chapter describes the technical and algorithmic evaluation of FINALLY that I designed and conducted. The evaluation examines whether the generated dataset sets satisfy the configured constraints and whether their diversity scores reflect the selected recommendation strategy across systematically varied configurations. It additionally assesses the reproducibility of the deterministic strategies and characterizes the end-to-end response times of the complete recommendation workflow.

Reproducible recommender-systems research requires transparent reporting of experimental settings, implementations, and evaluation artifacts \cite{beel2016reproducibility}. Prior benchmarking work has additionally shown that differences in evaluation choices can affect the reproducibility and comparability of recommendation results \cite{sun2020evaluating}. Recent methodological analyses therefore emphasize the explicit documentation of evaluation configurations, metrics, repetitions, and analysis procedures \cite{bauer2024evaluation,jannach2026methodological}.

The following sections define the evaluation objectives, experimental setup, evaluation metrics, and experimental procedure.

\section{Evaluation Objectives}
\label{sec:evaluation-objectives}

The evaluation primarily addresses the second research question by examining to what extent the dataset sets generated by FINALLY satisfy user-defined constraints and reflect the selected recommendation strategy across different configurations. Four evaluation objectives operationalize this question:

\begin{enumerate}
    \item \textbf{Evaluation completeness and constraint compliance.}
    The evaluation examines whether all planned runs were completed and whether the generated dataset sets satisfy the configured target-set size, metadata filters, required-dataset constraints, duplicate restrictions, and snapshot-membership requirements.

    \item \textbf{Strategy alignment.}
    The evaluation examines whether the \gls{aps}-based strategies produce diversity scores in the intended direction. For each diversity measure, the score of the diverse variant is compared with that of the corresponding non-diverse variant within the same configuration. The deterministic results are additionally positioned relative to configuration-specific Random selections.

    \item \textbf{Deterministic reproducibility.}
    The evaluation examines whether repeated executions of an unchanged deterministic strategy and configuration return the same dataset membership, recorded output order, and diversity scores.

    \item \textbf{End-to-end response time.}
    The evaluation measures the duration of the complete browser-based recommendation workflow and examines how the response times change across strategies and target-set sizes.
\end{enumerate}

The evaluation is technical and algorithmic and does not include a user study. Consequently, it does not assess perceived usability, researcher satisfaction, or the time required by researchers to construct dataset sets manually. The strategy-alignment analysis evaluates whether the generated dataset sets produce scores in the expected direction under the selected objective. It does not establish global optimality of the greedy multi-start procedure.

\section{Experimental Setup}
\label{sec:evaluation-setup}

The evaluation included the four deterministic strategies introduced in Chapter~\ref{chap:finally}: Diverse Convex Hull, Non-Diverse Convex Hull, Diverse \gls{effcov}, and Non-Diverse \gls{effcov}, together with Random selection.

Across these strategies, the selected performance metric, cutoff, target-set size, metadata filters, and the inclusion of a required dataset were varied. This resulted in the ten configurations shown in Table~\ref{tab:evaluation-configurations}.

\begin{table}[htbp]
    \centering
    \small
    \caption{Experimental configurations. Candidate-pool sizes exclude
    separately specified required datasets.}
    \label{tab:evaluation-configurations}
    \begin{tabularx}{\textwidth}{
    >{\raggedright\arraybackslash}p{5.5cm}
    >{\centering\arraybackslash}p{2.2cm}
    >{\raggedright\arraybackslash}X
    r
    }
        \toprule
        Configuration
        & Metadata filters
        & Required dataset
        & Candidate pool \\
        \midrule

        \texttt{ndcg10-size5}
        & No
        & None
        & 96 \\

        \texttt{ndcg10-size10}
        & No
        & None
        & 96 \\

        \texttt{ndcg10-size20}
        & No
        & None
        & 96 \\

        \texttt{hr10-size5}
        & No
        & None
        & 96 \\

        \texttt{recall10-size5}
        & No
        & None
        & 96 \\

        \texttt{ndcg1-size5}
        & No
        & None
        & 96 \\

        \texttt{ndcg20-size5}
        & No
        & None
        & 96 \\

        \texttt{filtered-ndcg10-size5}
        & Yes
        & None
        & 13 \\

        \texttt{required-ndcg10-size5}
        & No
        & \texttt{Amazon2014-Video-Games}
        & 95 \\

        \texttt{filtered-required-ndcg10-size5}
        & Yes
        & \texttt{Amazon2014-Video-Games}
        & 13 \\

        \bottomrule
    \end{tabularx}
\end{table}

I designed the experiment around \texttt{ndcg10-size5} as a baseline configuration. \gls{ndcg} at cutoff 10 was used as a fixed reference point from which the performance metric and cutoff could be varied independently: the metric variations retained cutoff 10, whereas the cutoff variations retained \gls{ndcg}. The choice was intended to provide a consistent reference configuration rather than to imply that \gls{ndcg}@10 is generally preferable to the other supported metric--cutoff combinations. A target-set size of five provided the initial selection problem and was subsequently doubled and quadrupled to examine increasingly larger selections.

Six configurations varied one factor of this baseline at a time: the target-set size was increased from 5 to 10 and 20 datasets, the performance metric was changed from \gls{ndcg} to \gls{hr} and Recall while retaining cutoff 10, and the \gls{ndcg} cutoff was changed from 10 to 1 and 20. Two further configurations independently introduced metadata filtering and a required dataset, while the final configuration combined both mechanisms. The ten configurations therefore cover the baseline, systematic single-factor variations, and one combined constraint case without attempting to exhaust the complete configuration space supported by FINALLY.

The configuration labels in this thesis\footnote{The evaluation repository preserves several historical identifiers, including \texttt{baseline-ndcg10-size5}, \texttt{size10-ndcg10}, \texttt{size20-ndcg10}, \texttt{metric-hr10-size5}, \texttt{metric-recall10-size5}, \texttt{cutoff-ndcg1-size5}, and \texttt{cutoff-ndcg20-size5}. The remaining identifiers are identical to the labels used in this thesis.} follow the pattern:
\begin{center}
    \small
    \texttt{<optional-prefix>-<metric><cutoff>-size<target>}
\end{center}
The metric and cutoff specify the performance measure used to construct the \gls{aps} performance vectors, while the final component denotes the requested number of datasets. For example, \texttt{ndcg10-size5} denotes the use of \gls{ndcg} at cutoff 10 with a target-set size of five datasets, whereas \texttt{hr10-size5} uses Hit Rate at cutoff 10. The prefixes \texttt{filtered-} and \texttt{required-} indicate the activation of metadata filters and the inclusion of a required dataset, respectively. Both prefixes are combined in \texttt{filtered-required-ndcg10-size5}.

The target-set sizes 5, 10, and 20 provide progressively larger selection problems while retaining the same candidate pool in the corresponding unfiltered configurations. A target-set size of five serves as the baseline, while sizes 10 and 20 double and quadruple this value. This variation allows the evaluation to examine whether directional strategy behavior remains consistent and how end-to-end response time changes as the number of datasets to be selected increases. The seven unfiltered configurations without a required dataset used all 96 datasets contained in the evaluation snapshot.

The metadata thresholds were derived from the distribution of dataset characteristics in the evaluation snapshot rather than chosen as arbitrary absolute values. The interaction threshold of 236,359 corresponds to the median number of interactions, while the originally requested density threshold of 0.182996 corresponds to the median dataset density. Applying both lower bounds simultaneously reduced the eligible candidate pool from 96 to 13 datasets while retaining enough candidates to construct the requested target set. This filtered configuration was included to examine whether constraint compliance and strategy alignment were preserved when the recommendation procedure operated on a substantially restricted search space rather than the complete dataset pool. Because FINALLY stores and represents dataset density in percentage points, the requested threshold of 0.182996 corresponds to 0.182996\% density. The user interface represents density values with two decimal places, so the effective density threshold applied during recommendation generation was 0.18\%.

The required-dataset configurations used \texttt{Amazon2014-Video-Games}. During the configuration design, this dataset was identified as the dataset whose number of interactions, 231,780, was closest to the median of 236,359 while retaining usable performance values for all three evaluated metrics at cutoff 10. In \texttt{required-ndcg10-size5}, it was fixed in the final set and removed from the candidate pool, leaving 95 candidates. The dataset also falls below both thresholds used in the combined filtered configuration. Consequently, \texttt{filtered-required-ndcg10-size5} explicitly exercises the intended distinction between candidate filters and required datasets: candidate datasets must satisfy the active filters, whereas the separately required dataset remains part of the final set independently of them.

Each of the four deterministic strategies was executed three times per configuration. These repetitions were used to examine whether unchanged strategy--configuration combinations produced identical dataset membership, output order, and diversity scores. Random selection was executed 30 times per configuration to obtain a configuration-specific empirical reference distribution for both diversity objectives. The complete experiment therefore comprised 420 recommendation runs:

\[
    10 \times \left(4 \times 3 + 30\right) = 420.
\]

The offline analysis used an immutable snapshot retrieved on 26 July 2026, at 20:23 UTC. The snapshot contained 96 datasets, 29 algorithms, and 2,676 recorded dataset--algorithm performance pairs out of 2,784 possible pairs. The remaining 108 pairs were absent. It included \gls{ndcg}, \gls{hr}, and Recall values at cutoffs 1, 3, 5, 10, and 20.

Missing dataset--algorithm dimensions were imputed using the corresponding column mean calculated over the active evaluation domain. If an algorithm dimension had been entirely absent from that domain, it would have been replaced with zero.

The browser-based experiment collector interacted with the deployed FINALLY instance at \url{https://finally.recommender-systems.com/}. It was executed on a Windows 11 x64 system equipped with an Intel Core i5-13600K processor and 32~GB of DDR4 memory, using Node.js 24.17.0, TypeScript 5.9.3, Playwright 1.62.0, and Chromium 151.0.7922.34. The system accessed the deployed application through a wireless network connection. The same client hardware and software environment were used throughout the final collection.

The final evaluation used FINALLY commit~\texttt{48943683}\footnote{Full commit hash: \nolinkurl{489436834f95300c84bcaf3843b3c472eb1e8be3}.}. Before collection, nine relevant browser assets served by the deployed FINALLY instance were compared with the corresponding files from this source state and matched after canonicalizing line endings. The evaluation therefore operated against a verified deployment of the implementation described in Chapter~\ref{chap:finally}. The immutable snapshot and the evaluated FINALLY source state were treated as read-only inputs during collection.

The complete final evaluation data, configuration files, collector and analysis source code, raw run records, processed results, and reproduction instructions are archived in the fixed evaluation repository release\footnote{\url{https://github.com/LouisOwie/FINALLY_evaluation/releases/tag/v2.0.0}}. The final experiment is identified as \texttt{final-corrected-v1}; the earlier preliminary experiment is retained separately in the repository for provenance but is not used as evidence for the results reported in this thesis.

\section{Evaluation Metrics}
\label{sec:evaluation-metrics}

The evaluation metrics were organized around the four objectives introduced in Section~\ref{sec:evaluation-objectives}: evaluation completeness and constraint compliance, strategy alignment, deterministic reproducibility, and end-to-end response time.

\paragraph{Evaluation completeness and constraint compliance.}

A planned run was counted as successfully completed if the browser-based workflow reached a terminal success state and a valid result record was stored for its unique run identifier. The evaluation matrix was considered complete if a successful final record existed for each of the 420 planned run identifiers.

Constraint compliance was evaluated using five criteria. First, target-size compliance required the number of datasets in the final set to equal the configured target-set size. Second, duplicate-free compliance required every dataset name within a final set to be unique. Third, snapshot membership required every returned dataset to be contained in the immutable evaluation snapshot. Fourth, required-dataset inclusion required all separately specified required datasets to be present in the final set. This criterion was applied only to configurations containing a required dataset.

Finally, metadata-filter compliance examined whether all newly recommended datasets satisfied each configured filter. Required datasets were excluded from this check because, as described in Section~\ref{sec:evaluation-setup}, they were retained independently of the candidate filters. Two configurations contained two active metadata filters each. Since every configuration comprised 42 runs, the total number of configured filter checks was

\[
    2 \text{ configurations}
    \times 42 \text{ runs}
    \times 2 \text{ filters}
    = 168.
\]
Each check assessed whether all recommended datasets within the corresponding
run satisfied the respective filter.

\paragraph{Score reconstruction and strategy alignment.}

The Convex Hull and \gls{effcov} scores of every collected final dataset set were reconstructed offline from the immutable evaluation snapshot. The reconstruction used the exact metric, cutoff, candidate pool, required datasets, and final dataset membership recorded for the corresponding run. This information allowed the same performance-vector construction and imputation domain used during recommendation generation to be reproduced independently of the deployed application.

For the offline analysis, I implemented the diversity calculations separately from the FINALLY application. Before the final evaluation, automated tests compared its \gls{mnnd}, intrinsic Convex Hull, Convex Hull diversity, and \gls{effcov} calculations with the corresponding FINALLY implementation on representative numerical fixtures. Differences were accepted only within a mixed absolute-relative tolerance of \(10^{-12}\). For every final run, both diversity objectives were reconstructed, resulting in
\[
    420 \times 2 = 840
\]
offline score values. As an additional validation step, these values were compared with direct evaluations of the corresponding FINALLY metric functions over the exact stored imputation domain and final dataset set of each run. The result of this comparison is reported in Section~\ref{sec:results-constraint-compliance}.

Strategy alignment was evaluated separately for the two diversity objectives. Each strategy was evaluated using its corresponding objective: the Diverse and Non-Diverse Convex Hull strategies were compared using the Convex Hull score, whereas the Diverse and Non-Diverse \gls{effcov} strategies were compared using the \gls{effcov} score.

For a given objective \(o\) and configuration \(c\), let \(s_{\mathrm{diverse}}^{(o,c)}\) denote the diversity score produced by the Diverse strategy and \(s_{\mathrm{non\text{-}diverse}}^{(o,c)}\) the score produced by the corresponding Non-Diverse strategy. The expected directional ordering was

\[
    s_{\mathrm{diverse}}^{(o,c)}
    >
    s_{\mathrm{non\text{-}diverse}}^{(o,c)}.
\]

A tie therefore did not satisfy the expected pairwise ordering. This comparison evaluates whether the generated sets produce scores in the direction intended by the selected strategy. It does not determine whether either strategy found a globally optimal dataset set.

Scores were compared only within the same configuration. Changes in the selected metric, cutoff, target-set size, candidate pool, or required datasets alter the underlying performance vectors and score distributions. Scores from different configurations therefore do not provide a common numerical scale.

The deterministic results were additionally positioned relative to the 30
Random runs belonging to the same configuration. Let \(R = \{r_1, \ldots, r_{30}\}\) denote the multiset of Random scores for the same configuration and diversity objective, and let \(s\) denote the deterministic diversity score being evaluated. The empirical percentile rank, hereafter referred to as the percentile, is defined as

\[
    P(s)
    =
    \frac{
        \left|
            \left\{
                r \in R \mid r \leq s
            \right\}
        \right|
    }{
        |R|
    }
    \times 100.
\]

The percentile \(P(s)\) therefore represents the percentage of Random scores that were less than or equal to the evaluated deterministic score. Random scores equal to \(s\) are included because the numerator uses \(r \leq s\). Consequently, \(P(s)=100\%\) means that no Random score was greater than \(s\), but strict superiority can only be established by additionally checking whether any equal Random scores occurred.

\paragraph{Deterministic reproducibility.}

Deterministic reproducibility was assessed across the three repetitions of each deterministic strategy--configuration combination. A combination was considered reproducible if all repetitions returned the same dataset membership, preserved the same recorded output order, and produced matching reconstructed Convex Hull and \gls{effcov} scores. Two finite score values \(s_1\) and \(s_2\) were considered numerically equal if
\[
    |s_1-s_2|
    \leq
    10^{-12}
    \max\left(
        1,
        |s_1|,
        |s_2|
    \right).
\]
Dataset membership and recorded output order were compared exactly.

Response times were not included in the reproducibility criterion because they are affected by system load, browser execution, and network conditions even when the generated dataset set remains unchanged.

\paragraph{End-to-end response time.}

End-to-end response time was measured from immediately before the browser triggered the Generate action until FINALLY reached a terminal success state confirming that the requested final dataset set had been generated. The subsequent extraction and offline analysis of the result records were not part of the measured interval.

The measurement therefore includes the browser interaction, communication with the deployed FINALLY instance, recommendation generation, and the corresponding user-interface status update. It represents a system-level response time and not an isolated benchmark of the Convex Hull, \gls{effcov}, or greedy selection implementations.

Response times were evaluated descriptively for every configuration--strategy combination. The analysis reported the number of observations \(n\), arithmetic mean, population standard deviation, minimum, and maximum. The deterministic strategies therefore contributed three observations per configuration and Random selection 30 observations. No inferential runtime comparison was performed.

\section{Experimental Procedure}
\label{sec:evaluation-procedure}

I conducted the final evaluation in two principal stages: browser-based recommendation collection and offline validation and aggregation. Before collection, I performed a preflight check of the experiment definition, snapshot, software environment, deployed FINALLY assets, and evaluation tests. The numerical diagnostics associated with the preliminary evaluation belonged to the preceding implementation-refinement process described in Section~\ref{sec:iterative-development} and were not part of the final evaluation analyzed in this chapter.

\paragraph{Browser-based recommendation collection.}

Before the main experiment, the evaluation configuration and the immutable snapshot were validated to ensure that all referenced configurations, datasets, metrics, cutoffs, and candidate-pool sizes were available. The experiment manifest assigned a unique run identifier to each of the 420 planned runs.

The complete set of 420 jobs was expanded before execution and then reordered through a seeded Fisher--Yates shuffle using the fixed runner seed \(20260726\). The resulting order was stored in the experiment manifest and executed sequentially without regrouping jobs by configuration or strategy. A fresh isolated browser context was created for each job. No warm-up runs were performed, and the final execution order matched the order recorded in the manifest.

For each run, the browser-based collector opened the deployed FINALLY application and applied the strategy, metric, cutoff, target-set size, metadata filters, and required datasets specified by the corresponding configuration. The collector subsequently read the effective filter values back from the user interface. This distinction between requested and applied values was necessary for the density filter, for which the requested value of 0.182996 was represented and applied by the interface as 0.18.

The collector also recorded the exact candidate pool associated with the run. The pool was derived from the datasets selected in the interface and matched against the immutable snapshot. Required datasets were removed from the candidate pool because they were handled separately, while datasets without finite performance values for the selected metric and cutoff were excluded from recommendation.

After the configuration had been applied, the collector triggered the recommendation and waited until FINALLY reported a terminal success state. The elapsed time between triggering the recommendation and reaching this state was stored as the end-to-end response time defined in Section~\ref{sec:evaluation-metrics}. The generated result table was then parsed in its displayed order.

Before a run was stored, the collector verified that the result contained the requested number of datasets, that all dataset names and status values were present, that no duplicate dataset names occurred, that all configured required datasets were included, and that every returned dataset belonged to the corresponding candidate pool or was explicitly marked as required.

Each successfully collected run was stored as an individual raw \gls{json} record. In addition to the returned dataset set, the record contained its unique run identifier, configuration, strategy, repetition number, requested and applied filters, required and recommended datasets, candidate pool, recorded output order, response time, timestamp, and source \gls{url}. Existing raw records were not overwritten during collection.

\paragraph{Offline validation and aggregation.}

The raw run records were subsequently processed against the immutable evaluation snapshot. The datasets listed as required or recommended in the result table formed the final dataset set while preserving their recorded order. The structural and metadata checks defined in Section~\ref{sec:evaluation-metrics} were then applied to each processed run.

For score reconstruction, the imputation domain consisted of the exact candidate pool recorded for the run together with its required datasets. The algorithms were ordered consistently by name, and the selected performance metric and cutoff were extracted for every available dataset--algorithm pair. If multiple performance records existed for the same pair, their finite values were averaged. Datasets without any finite value for the selected metric and cutoff were excluded from the domain. Remaining missing dimensions were imputed as described in Section~\ref{sec:evaluation-setup}.

The performance vectors corresponding to the final dataset set were then selected from this domain, and both the Convex Hull and \gls{effcov} scores were reconstructed using the independently maintained offline metric implementation described in Section~\ref{sec:evaluation-metrics}. The processed records were subsequently grouped by configuration and strategy to derive the constraint summaries, strategy-alignment comparisons, Random percentiles, deterministic reproducibility checks, and descriptive response-time statistics.

Completeness was assessed by comparing the unique run identifiers of the final raw records with the 420 identifiers defined in the experiment manifest. Missing identifiers, duplicate final records, and jobs without a successful final record were reported separately.

\paragraph{Reproducibility artifacts.}

Release~\texttt{v2.0.0} of the evaluation repository contains the final experiment definition, execution manifest, browser-based collector, immutable snapshot, raw records for all 420 final runs, offline validation and aggregation code, processed result files, source-state provenance, and reproduction instructions. The archived snapshot and raw records are sufficient to reproduce the offline analysis without executing new recommendations against the deployed FINALLY application. The preliminary experiment and its diagnostic artifacts are retained in separate directories and are not included in the final result analysis.

% ============================================================================
% CHAPTER 6: RESULTS
%
% Present empirical observations as objectively as possible.
%
% Report:
%
% - exact values,
% - effect sizes,
% - uncertainty,
% - statistical tests where applicable, and
% - relevant differences between methods or conditions.
%
% Do not merely write "Method A performs better than Method B". State the
% values and quantify the difference.
%
% Avoid presenting many nearly identical charts. Select a visualisation that
% communicates the central result without unnecessary repetition.
%
% Reserve interpretation primarily for the discussion chapter unless results
% and discussion are intentionally combined.
% ============================================================================

\cleardoublepage

\chapter{Results}
\label{chap:results}

This chapter presents the results of the technical and algorithmic evaluation of FINALLY conducted in this thesis. First, the completeness of the collected evaluation data and the compliance of the generated dataset sets with the configured constraints are examined. Subsequently, the alignment between the selected recommendation strategies and the resulting dataset sets is analyzed. The chapter concludes with results on deterministic reproducibility and end-to-end response times.

\section{Evaluation Completeness and Constraint \break Compliance}
\label{sec:results-constraint-compliance}

The final evaluation matrix comprised 420 successful runs across ten configurations. All 420 planned jobs succeeded on their first attempt and no final failed attempts, missing run identifiers, or duplicate final records occurred. Table~\ref{tab:constraint-compliance} summarizes the completeness and constraint-compliance checks.

\begin{table}[htbp]
    \centering
    \caption{Evaluation-completeness and constraint-compliance results.}
    \label{tab:constraint-compliance}
    \begin{tabular}{lrr}
        \toprule
        Criterion & Satisfied & Compliance \\
        \midrule
        Successful final runs                  & 420 / 420 & 100\% \\
        Requested target-set size              & 420 / 420 & 100\% \\
        Duplicate-free dataset sets            & 420 / 420 & 100\% \\
        Required-dataset inclusion             & 84 / 84   & 100\% \\
        Runs containing only snapshot datasets & 420 / 420 & 100\% \\
        Configured metadata-filter checks      & 168 / 168 & 100\% \\
        Offline/direct diversity-score agreement & 840 / 840 & 100\% \\
        \bottomrule
    \end{tabular}
\end{table}

Two configurations specified a required dataset: \texttt{required-ndcg10-size5} and \texttt{filtered-required-ndcg10-size5}. Together, these configurations accounted for 84 runs. The required dataset was included in the final dataset set in all 84 applicable runs. The remaining configurations did not specify a required dataset and were therefore not included in this denominator.

The 168 configured metadata-filter checks defined in Section~\ref{sec:evaluation-metrics} were all satisfied. No configured filter was violated, and no check was classified as non-evaluable. The interaction threshold was applied as requested at 236,359 interactions. Density compliance refers to the effective user-interface threshold of 0.18 rather than the originally requested value of 0.182996. Required datasets were assessed separately from newly recommended datasets because they were retained independently of the active candidate filters.

Finally, both diversity objectives were reconstructed offline for every final dataset set, resulting in 840 score values across the 420 runs. All 840 values matched the corresponding values obtained by directly evaluating FINALLY's metric functions over the same stored domains and final sets. The maximum absolute difference and maximum relative difference were both zero. Thus, no run had to be excluded from the strategy-alignment analysis because of missing datasets, malformed results, constraint violations, or unavailable diversity scores.

Overall, the generated dataset sets satisfied all structural, required-dataset, and metadata-filter constraints covered by the experimental configurations. These results establish the completeness and internal consistency of the collected run set but do not yet determine whether the selected datasets reflect the intended diverse or non-diverse recommendation objectives. This is examined in the following section.

\section{Strategy Alignment}
\label{sec:results-strategy-alignment}

Using the directional-alignment criterion and the configuration-specific Random percentiles defined in Section~\ref{sec:evaluation-metrics}, the analysis compared each \gls{aps}-based strategy with its corresponding counterpart and with the 30 Random runs under the same configuration. All comparisons were conducted within configurations, as the underlying score scales differ across metrics, cutoffs, target-set sizes, candidate pools, and required datasets.

\paragraph{\gls{effcov}-Based Selection.}
\label{sec:results-effcov}

The \gls{effcov}-based strategies showed consistent directional alignment across all ten configurations. In every configuration, the dataset set generated by Diverse \gls{effcov} obtained a higher \gls{effcov} score than the set generated by Non-Diverse \gls{effcov}. Thus, the expected pairwise score direction was observed in all ten configurations.

In \texttt{ndcg10-size5}, Diverse \gls{effcov} produced a score of 0.565790, whereas Non-Diverse \gls{effcov} produced a score of approximately \(2.31 \times 10^{-5}\). In the configuration \texttt{filtered-required-ndcg10-size5}, the corresponding scores were 0.106335 and approximately \(3.71 \times 10^{-4}\). The expected ordering also held across all investigated metric, cutoff, target-size, filter, and required-dataset variations.

\paragraph{Convex-Hull-Based Selection.}
\label{sec:results-convex-hull}

The Convex-Hull-based strategies showed the expected pairwise score direction across all ten experimental configurations. In every configuration, the dataset set generated by Diverse Convex Hull obtained a higher Convex Hull score than the corresponding set generated by Non-Diverse Convex Hull. Table~\ref{tab:convex-hull-alignment} summarizes the resulting alignment.

\begin{table}[htbp]
    \centering
    \caption{Pairwise strategy-alignment results for Convex-Hull-based selection.}
    \label{tab:convex-hull-alignment}
    \begin{tabular}{lr}
        \toprule
        Result & Configurations \\
        \midrule
        Diverse score $>$ Non-Diverse score & 10 / 10 \\
        Diverse score $<$ Non-Diverse score & 0 / 10 \\
        Equal scores                         & 0 / 10 \\
        \bottomrule
    \end{tabular}
\end{table}

For \texttt{hr10-size5}, Diverse Convex Hull produced a score of 0.592497, whereas Non-Diverse Convex Hull produced approximately \(1.68 \times 10^{-8}\). The strict ordering therefore also held for the \gls{hr}@10 performance space.

The smallest absolute Convex Hull scores occurred for the larger target sets. In \texttt{ndcg10-size10}, Diverse Convex Hull produced approximately \(1.26 \times 10^{-9}\), compared with approximately \(1.01 \times 10^{-28}\) for Non-Diverse Convex Hull. In \texttt{ndcg10-size20}, the corresponding scores were approximately \(3.57 \times 10^{-32}\) and \(3.08 \times 10^{-64}\). Despite their small absolute magnitudes, these values remained distinct and preserved the intended maximizing and minimizing directions.

Overall, Convex-Hull-based selection satisfied the directional-alignment criterion in all evaluated configurations. As with \gls{effcov}, this result establishes alignment with the implemented objective but does not demonstrate global optimality of the selected dataset sets.

\paragraph{Comparison with Random Selection.}
\label{sec:results-random-comparison}

The 30 Random runs per configuration provided empirical reference distributions for both diversity objectives. Across the final evaluation, the deterministic strategies showed complete directional separation from these configuration-specific Random distributions. Table~\ref{tab:random-comparison} summarizes the results for the four deterministic strategies under their corresponding selection objectives.

\begin{table}[htbp]
    \centering
    \caption{Position of the deterministic strategies relative to the configuration-specific Random score distributions under their corresponding objectives.}
    \label{tab:random-comparison}
    \begin{tabular}{lp{8.3cm}}
        \toprule
        Strategy & Result relative to Random selection \\
        \midrule
        Diverse \gls{effcov}
            & Above all 30 Random \gls{effcov} scores in all 10 configurations. \\

        Non-Diverse \gls{effcov}
            & Below all 30 Random \gls{effcov} scores in all 10 configurations. \\

        Diverse Convex Hull
            & Above all 30 Random Convex Hull scores in all 10 configurations. \\

        Non-Diverse Convex Hull
            & Below all 30 Random Convex Hull scores in all 10 configurations. \\
        \bottomrule
    \end{tabular}
\end{table}

For the diverse strategies, the corresponding objective score exceeded the maximum Random score in every configuration. Their empirical percentile was therefore \(100\%\) in all 20 strategy--configuration comparisons. Conversely, both non-diverse strategies produced objective scores below the minimum Random score in every configuration and therefore obtained an empirical percentile of \(0\%\) in all 20 corresponding comparisons. No exact ties with Random scores occurred.

The broader cross-objective analysis produced the same directional pattern. Each of the 40 deterministic results was evaluated under both diversity objectives, resulting in 80 deterministic-result--objective comparisons. All 40 comparisons associated with a diverse strategy were above the complete corresponding Random distribution, whereas all 40 comparisons associated with a non-diverse strategy were below it. No exact Random ties occurred in any of these comparisons.

The Random reference therefore supports the pairwise strategy-alignment results for both objectives: within every evaluated configuration, the maximizing strategies produced sets at the upper extreme of the observed Random score distribution, whereas the minimizing strategies produced sets at the lower extreme. These empirical comparisons describe the position of the deterministic results relative to 30 sampled Random sets and do not imply that the deterministic solutions are global optima.

\section{Deterministic Reproducibility}
\label{sec:results-reproducibility}

The reproducibility analysis covered the four deterministic recommendation strategies across all ten configurations. Each combination of strategy and configuration was executed three times, resulting in 120 deterministic runs and 40 distinct combinations.

All three repetitions returned the same dataset membership and the same recorded output order for every combination of strategy and configuration. Membership reproducibility and recorded-order reproducibility were therefore observed in 40 of 40 deterministic combinations. The reconstructed Convex Hull scores and \gls{effcov} scores also satisfied the reproducibility criterion in all 40 combinations.

Random selection was excluded from this comparison because its output is intentionally stochastic. Its 30 repetitions were instead used to construct the empirical reference distributions reported in Section~\ref{sec:results-random-comparison}.

Overall, all 40 deterministic strategy--configuration combinations were reproducible with respect to dataset membership, recorded output order, Convex Hull score, and \gls{effcov} score. Thus, no non-deterministic variation in the output of the four \gls{aps}-based selection strategies was observed across the three repetitions.

\section{End-to-End Response Times}
\label{sec:results-response-times}

Using the end-to-end response-time definition from Section~\ref{sec:evaluation-metrics}, the analysis descriptively compared the observed system-level durations across strategies and configurations. The clearest differences occurred when increasing the requested target-set size. Table~\ref{tab:response-times-size} reports the corresponding mean response times for \texttt{ndcg10-size5}, \texttt{ndcg10-size10}, and \texttt{ndcg10-size20}.

Table~\ref{tab:response-times-size} focuses on the target-size variation to illustrate the main response-time pattern without overloading the presentation. Complete descriptive response-time statistics for all 50 configuration--strategy combinations, including the number of observations, mean, population standard deviation, minimum, and maximum, are provided in Appendix~\ref{app:response-time-statistics}.

\begin{table}[htbp]
    \centering
    \small
    \caption{Mean end-to-end response times in seconds for increasing target-set sizes. Deterministic strategies were repeated three times and Random selection 30 times.}
    \label{tab:response-times-size}
    \begin{tabular}{@{}lccccc@{}}
        \toprule
        Configuration
        & Div. CH
        & Non-Div. CH
        & Div. \gls{effcov}
        & Non-Div. \gls{effcov}
        & Random \\
        \midrule

        \texttt{ndcg10-size5}
        & 0.277
        & 0.279
        & 0.236
        & 0.235
        & 0.153 \\

        \texttt{ndcg10-size10}
        & 1.200
        & 1.214
        & 0.997
        & 1.010
        & 0.151 \\

        \texttt{ndcg10-size20}
        & 9.434
        & 9.589
        & 8.326
        & 8.516
        & 0.151 \\

        \bottomrule
    \end{tabular}

    \vspace{0.4em}
    \parbox{\textwidth}{\footnotesize\textit{Notes:} CH denotes Convex Hull. Div. and Non-Div. denote the diverse and non-diverse strategy variants.}
\end{table}

For a target-set size of five, the Convex Hull strategies required mean response times of 0.277 seconds for the diverse variant and 0.279 seconds for the non-diverse variant. The corresponding \gls{effcov} means were 0.236 and 0.235 seconds. Random selection required a mean of 0.153 seconds.

Increasing the target-set size to ten raised the mean response times of the Convex Hull strategies to 1.200 and 1.214 seconds. Diverse and Non-Diverse \gls{effcov} required 0.997 and 1.010 seconds, respectively, whereas the Random mean remained at approximately 0.151 seconds.

The largest response times among the target-size configurations occurred for \texttt{ndcg10-size20}. Diverse Convex Hull required a mean of 9.434 seconds and Non-Diverse Convex Hull 9.589 seconds. The corresponding \gls{effcov} means were 8.326 and 8.516 seconds. Random selection remained at approximately 0.151 seconds. Thus, the response times of both \gls{aps}-based objectives increased substantially with the number of datasets to be selected, while the increase was larger for Convex Hull than for \gls{effcov}.

The remaining unfiltered configurations with a target-set size of five showed similar response times. Across the metric- and cutoff-variation configurations, mean Convex Hull response times ranged from approximately 0.276 to 0.280 seconds, \gls{effcov} means from approximately 0.235 to 0.241 seconds, and Random means from approximately 0.148 to 0.152 seconds.

The filtered configurations differed from this pattern. In \texttt{filtered-ndcg10-size5}, mean response times ranged from 0.491 to 0.499 seconds across the five strategies, while \texttt{filtered-required-ndcg10-size5} ranged from 0.484 to 0.491 seconds. By contrast, \texttt{required-ndcg10-size5} produced means between approximately 0.143 and 0.148 seconds. Since the measured interval covers the complete browser-based recommendation workflow, these differences cannot be attributed solely to the numerical selection procedures.

Overall, Random selection remained close to 0.15 seconds across the unfiltered target-size configurations, whereas the \gls{aps}-based strategies became progressively slower as the requested set size increased. At target-set size 20, the \gls{effcov} variants required approximately 88--89\% of the response time of their corresponding Convex Hull variants. These measurements are descriptive system-level results and do not isolate algorithm execution time from browser, network, server, or interface effects.

% ============================================================================
% CHAPTER 7: DISCUSSION
%
% Interpret the reported results and explain their scientific or practical
% meaning.
%
% Explicitly:
%
% - answer the research questions,
% - relate the findings to previous work,
% - explain unexpected results,
% - distinguish statistical from practical relevance,
% - discuss alternative explanations, and
% - identify threats to validity.
%
% The reader should not have to infer the central conclusion from a table or
% figure without guidance.
% ============================================================================

\cleardoublepage

\chapter{Discussion}

\label{chap:discussion}

Chapter~\ref{chap:results} reported the results of the technical and algorithmic evaluation of \mbox{FINALLY}. This chapter interprets their scientific and practical meaning, relates the findings to previous work, and discusses the implications of the observed strategy behavior and response times. It subsequently examines the threats to the validity of the evaluation before deriving explicit answers to the two research questions.

\section{Interpretation and Relation to Previous Work}
\label{sec:discussion-interpretation}

The complete constraint compliance reported in Section~\ref{sec:results-constraint-compliance} indicates that FINALLY consistently enforced the evaluated structural and metadata requirements. In particular, target-set size, duplicate prevention, required-dataset inclusion, and candidate filtering remained compatible across the investigated configurations. This finding supports implementation conformance within the evaluated configuration space, but not the scientific suitability of every generated dataset set. A structurally valid set may still be inappropriate for a particular research question, domain, or experimental protocol. Assessing this substantive suitability therefore remains the responsibility of the researcher.

Previous work has shown that dataset characteristics can affect the performance observed for recommender-system algorithms \cite{adomavicius2012data}. Chin et al. additionally demonstrate that recommendation datasets differ substantially and that their properties and usage are not always sufficiently understood \cite{chin2022datasets}. Against this background, the present results indicate that FINALLY can operationalize configurable \gls{aps}-based dataset-set construction by converting explicit user-defined constraints and selection objectives into complete dataset sets.

The FINALLY system developed in this thesis extends the line of work established by \gls{aps} and APS Explorer \cite{beel2024aps,vente2025apsexplorer}. Whereas these previous works provide the conceptual and interactive basis for performance-oriented dataset exploration, FINALLY applies this basis to the recommendation of configurable dataset sets. The evaluation does not establish that this workflow improves research decisions compared with manual selection, but it demonstrates that the underlying concepts can be integrated into a technically consistent selection process.

The 10-of-10 directional alignment for both implemented \gls{aps}-based objectives indicates that the diverse and non-diverse strategies consistently influenced the resulting dataset sets in their intended optimization directions within the investigated configuration space. The configuration-specific Random comparisons strengthen this interpretation: the deterministic diverse and non-diverse results consistently occupied opposite extremes of the sampled Random reference distributions under their corresponding objectives.

In relation to the \gls{aps}-based selection procedures described by Reising \cite{reising2026systematic}, these findings show that FINALLY consistently operationalized the intended maximizing and minimizing behavior under the evaluated conditions. However, FINALLY implements adapted forms of these objectives: the scores are not normalized by a set-size-specific reference value and omit the additive stabilizer \(\varepsilon\). The results therefore validate the behavior of the implemented FINALLY objectives rather than constituting an exact empirical reproduction of Reising's complete metric definitions.

Objective alignment must also be distinguished from scientific usefulness. The results show that the generated sets were more or less diverse according to the selected \gls{aps}-based objective, but they do not demonstrate that a highly diverse set necessarily produces a more informative recommender-systems evaluation. No downstream experiments examined whether the selected dataset sets reveal additional algorithmic strengths, weaknesses, or ranking changes. The findings should therefore be interpreted as evidence of technical consistency and objective alignment rather than as evidence that one selection strategy is scientifically superior to another.

\section{Strategy Behavior and Practical Implications}
\label{sec:discussion-implications}

The directional separation remained observable when the metric, cutoff, target-set size, candidate pool, and required-dataset setting were varied. This suggests that these configuration mechanisms can be combined without preventing the selected strategy from influencing the resulting dataset set in its intended direction.

The absolute magnitude of a diversity score should nevertheless not be interpreted independently of its configuration. This is particularly visible for Convex Hull at larger target-set sizes, where valid scores reached very small numerical magnitudes. For \texttt{ndcg10-size20}, for example, Diverse Convex Hull produced approximately \(3.57 \times 10^{-32}\), whereas Non-Diverse Convex Hull produced approximately \(3.08 \times 10^{-64}\). These values remained distinguishable by the scale-aware score comparison described in Section~\ref{sec:finally-selection}, but their magnitudes are specific to the underlying performance space and raw, unnormalized objective. Comparisons of absolute diversity values across configurations would therefore be inappropriate.

All 40 deterministic strategy--configuration combinations were reproducible, further strengthening the traceability of the selection behavior. This is consistent with the broader requirement for transparent and reproducible recommender-systems evaluations \cite{beel2016reproducibility,jannach2026methodological}. Reproducibility, however, should not be confused with global optimality: the greedy multi-start procedure remains a heuristic and may consistently reproduce a solution that is not the globally optimal subset under the corresponding objective.

The response-time results show that the end-to-end duration of the interactive workflow increases with the requested target-set size for both \gls{aps}-based objectives, while Random selection remained comparatively stable. Convex Hull also showed somewhat higher response times than \gls{effcov} at larger target-set sizes. Because the measurements include browser execution, network communication, server interaction, and interface updates, they cannot be interpreted as isolated algorithm runtimes.

Two implementation choices additionally affect how the resulting performance spaces are interpreted. First, FINALLY does not standardize individual algorithm dimensions before selection. Algorithm dimensions with greater performance variation can therefore contribute more strongly to Euclidean distances, covariance, and the resulting diversity scores. Second, missing performance values are replaced using column-mean imputation. This preserves a complete performance representation but moves missing dimensions toward the average of the available observations and can consequently reduce differences between datasets. Both choices can influence which dataset sets optimize the implemented objectives.

Finally, omitting the additive stabilizer \(\varepsilon\) used in Reising's formulation has a conceptual consequence. If the structural component of either adapted FINALLY objective becomes zero, the complete score becomes zero regardless of the corresponding \gls{mnnd}. This behavior was not responsible for a failure of directional alignment in the final evaluation, but it remains a difference between FINALLY's implemented objective and the original metric formulation and should be considered when interpreting configurations with degenerate performance-vector geometry.

\section{Threats to Validity}
\label{sec:discussion-validity}

The interpretation of the evaluation results is subject to several threats to validity. Transparent reporting of evaluation choices and limitations is important for the comparability and interpretation of recommender-systems research \cite{bauer2024evaluation,jannach2026methodological}. For clarity, the threats are organized into construct, internal, conclusion, and external validity.

\paragraph{Construct validity.}

Strategy alignment was operationalized using the same Convex Hull and \gls{effcov} objectives that the corresponding strategies were designed to optimize. The evaluation therefore measures objective alignment rather than an independent notion of scientific usefulness. Constraint compliance similarly measures adherence to configured requirements rather than substantive dataset suitability.

Because the evaluation did not include downstream recommender-systems experiments or a user study, it cannot establish whether the generated sets improve experimental insight, usability, cognitive effort, time savings, or researcher decision quality. The end-to-end response time also represents the complete browser-based workflow rather than isolated algorithm runtime.

\paragraph{Internal validity.}

The browser-based experiment was conducted against the deployed FINALLY instance through a wireless network connection. Variations in browser execution, network conditions, client load, server load, and the deployed environment may therefore have affected the recorded response times. Using the same client hardware and software environment throughout the final collection reduced systematic differences, while the seeded randomization of the complete run order reduced systematic coupling between execution order and particular configurations or strategies. Nevertheless, the experiment did not provide the control of a local algorithm-level benchmark.

The final evaluation was conducted only after the numerical defects identified during the preliminary evaluation had been corrected and verified. The deployed browser assets relevant to the recommendation workflow were compared with the fixed FINALLY source state before collection. The complete 420-run experiment was then repeated using this corrected implementation, and only this repeated experiment contributes to the reported results. This separation reduces the risk that findings from the defective preliminary implementation influenced the final empirical conclusions.

All final dataset sets were additionally processed against an immutable evaluation snapshot, and both diversity objectives were reconstructed for every run. The resulting 840 offline score values matched the corresponding values obtained by directly evaluating FINALLY's metric functions over the same stored domains and final sets. This provides a consistency check between the collected results, the archived evaluation data, and the independently maintained offline analysis implementation, although it cannot rule out a conceptual error shared by both implementations or by the underlying objective definition.

\paragraph{Conclusion validity.}

The evaluation covered ten configurations, three repetitions for each deterministic combination, and 30 repetitions for each Random configuration. The deterministic repetitions were sufficient to identify whether the observed outputs changed across repeated executions, but they were not intended to estimate runtime distributions or support inferential statistical comparisons. No confidence intervals or significance tests were calculated, and the reported runtime differences should therefore be interpreted descriptively.

The Random percentiles were derived from 30 observed scores per configuration. They provide a configuration-specific empirical reference, but their precision remains limited by the finite number of Random repetitions. Although no exact ties occurred in the final evaluation, the percentile definition includes Random scores equal to the deterministic score. The Random comparison should therefore be interpreted as an empirical positioning relative to the sampled Random sets rather than as proof of optimality or a complete characterization of all possible random selections.

The evaluation compared completed dataset sets but did not establish that the greedy multi-start procedure found globally optimal solutions. Directional alignment only requires the maximizing strategy to produce a higher score than the corresponding minimizing strategy. It does not measure the distance from the global maximum or minimum of the combinatorial search space. Consequently, the complete 10-of-10 alignment observed for both objectives demonstrates consistent optimization direction under the evaluated conditions, but not optimal subset selection.

\paragraph{External validity.}

The results are based on one immutable dataset and algorithm snapshot and on a limited set of experimental configurations. Although the evaluation varied metrics, cutoffs, target-set sizes, filters, and required-dataset inclusion, it covered only a subset of the configuration space supported by FINALLY. In particular, it examined one metadata-filter combination and one specific required dataset.

The observed strategy alignment may differ for other dataset collections, algorithm portfolios, performance metrics, candidate-pool sizes, or \glspl{aps}. In particular, the geometry of the performance vectors determines which numerical Convex Hull path is exercised, and configurations outside the evaluated design may create geometric cases not represented by the final experiment. The conclusions therefore apply directly to the evaluated FINALLY version and snapshot and should not be generalized without further evidence to arbitrary performance spaces.

The preprocessing decisions also limit generalization. Mean imputation and the absence of feature-wise standardization affect the distances and covariance structure on which the \gls{aps}-based objectives operate. Other missing-data treatments or scaling procedures could therefore produce different selected dataset sets even when the same underlying performance data are used. Finally, because no researchers interacted with the system as part of the evaluation, the results cannot be generalized to actual user behavior, acceptance, or the effectiveness of FINALLY in practical research workflows.

\section{Answers to the Research Questions}
\label{sec:discussion-research-questions}

Based on the interpretation of the evaluation results and the identified threats to validity, the two research questions can be answered as follows.

\paragraph{Research Question 1.}

\textit{How can the construction of configurable dataset sets for offline recommender-systems evaluations be supported through an integrated, recommendation-centered workflow?}

\begin{table}[htbp]
    \centering
    \scriptsize
    \caption{Traceability from the derived system requirements to the corresponding FINALLY functions and verification.}
    \label{tab:rq1-traceability}
    \begin{tabularx}{\textwidth}{
        >{\raggedright\arraybackslash}p{3.2cm}
        >{\raggedright\arraybackslash}p{4.0cm}
        >{\raggedright\arraybackslash}p{2.7cm}
        >{\raggedright\arraybackslash}X
    }
        \toprule
        Requirement
        & FINALLY function
        & Described in
        & Verification \\
        \midrule

        Required datasets
        & Fixed datasets retained in the generated set
        & Sec.~\ref{sec:finally-workflow}, \ref{sec:finally-selection}
        & Required-dataset inclusion \\

        Candidate selection
        & Explicit restriction of the eligible candidate pool
        & Sec.~\ref{sec:finally-workflow}
        & Candidate-pool membership check \\

        Metadata filters
        & Metadata-based restriction of eligible candidates
        & Sec.~\ref{sec:finally-workflow}
        & Metadata-filter compliance \\

        Configurable set size
        & Validated total dataset count
        & Sec.~\ref{sec:finally-workflow}
        & Target-size compliance \\

        Alternative selection objectives
        & Random, Diverse/Non-Diverse EffCov, Diverse/Non-Diverse Convex Hull
        & Sec.~\ref{sec:finally-selection}
        & Strategy alignment and Random comparison \\

        Result inspection
        & Metadata table and \gls{aps} visualization
        & Sec.~\ref{sec:finally-results}
        & Functional workflow verification \\

        Documentation and reuse
        & Export and sharing functions
        & Sec.~\ref{sec:finally-results}
        & Functional workflow verification \\

        Administrative recording
        & Usage logging and protected administrative interface
        & Sec.~\ref{sec:finally-architecture}
        & Functional workflow verification \\

        \bottomrule
    \end{tabularx}
\end{table}

To address this research question, I developed FINALLY as an integrated workflow that combines required datasets, candidate-pool definition, metadata constraints, configurable target-set size, alternative selection objectives, result inspection, visualization, and export. FINALLY thereby supports the transition from exploring individual datasets to constructing and documenting a complete dataset set within a single executable workflow. The evaluation does not establish that this workflow is easier, faster, or more effective than manual selection. Such claims would require evidence from actual researchers.

Table~\ref{tab:rq1-traceability} summarizes how the requirements derived in Section~\ref{sec:iterative-development} are represented in the implemented artifact and how their realization was examined in this thesis.

\paragraph{Research Question 2.}

\textit{To what extent do the generated dataset sets satisfy user-defined constraints and reflect the selected recommendation strategy across different configurations?}

Within the investigated configuration space, FINALLY satisfied all evaluated user-defined constraints and consistently reflected the selected recommendation direction. Both the \gls{effcov}-based and Convex-Hull-based strategies produced the expected diverse--non-diverse ordering in all ten configurations, the deterministic diverse and non-diverse results lay on opposite sides of all corresponding sampled Random results, and all 40 deterministic strategy--configuration combinations were reproducible. These findings establish technical consistency and objective alignment for the evaluated FINALLY version and data snapshot, but not global optimality of the greedy selections or scientific suitability of every generated set.

% Answer every research question explicitly.
% Use the same identifiers introduced in the introduction.

% ============================================================================
% CHAPTER 8: CONCLUSION
%
% Answer the research questions directly and explain the "so what?" of the
% thesis.
%
% Summarise:
%
% - the problem,
% - the method,
% - the central empirical results,
% - the contribution, and
% - the implications.
%
% Do not introduce new experiments, evidence, or arguments in this chapter.
% ============================================================================

\chapter{Conclusion}
\label{chap:conclusion}

Dataset selection defines the empirical conditions under which recommender-system algorithms are evaluated and compared. This thesis addressed the need for an integrated, recommendation-centered workflow that supports researchers in constructing complete dataset sets while accounting for required datasets, dataset-level constraints, target-set sizes, and alternative selection objectives. It developed FINALLY for this purpose and evaluated whether its recommendations satisfy the configured constraints and reflect the selected recommendation strategies.

To answer the first research question, I designed and implemented FINALLY, a web-based dataset recommender that combines configuration, dataset-set recommendation, result inspection, documentation, and export within a single workflow. FINALLY operationalizes Random selection together with diverse and non-diverse variants of adapted \gls{effcov} and Convex Hull objectives based on Reising's existing dataset-set diversity measures. The researcher nevertheless retains responsibility for assessing whether a generated set is appropriate for the intended experiment.

To answer the second research question, I evaluated FINALLY across 420 recommendation runs and ten systematically varied configurations. All evaluated dataset sets satisfied the applicable target-size, duplicate-avoidance, snapshot-membership, required-dataset, and metadata-filter requirements, and all 40 deterministic strategy--configuration combinations were reproducible. Both the \gls{effcov}-based and Convex-Hull-based strategies produced the expected diverse-versus-non-diverse score ordering in all ten configurations. Under their corresponding objectives, the diverse strategies additionally produced scores above all 30 configuration-specific Random results, whereas the non-diverse strategies produced scores below all 30 Random results.

Within the investigated technical scope and configuration space, FINALLY therefore met the research goal by providing an integrated workflow whose recommendations satisfied the evaluated constraints and consistently reflected the selected recommendation direction. The findings establish technical consistency and objective alignment for the evaluated FINALLY version and data snapshot, but not the scientific suitability, global optimality, or practical superiority of the generated selections.

% ============================================================================
% CHAPTER 9: FUTURE WORK AND LIMITATIONS
%
% State limitations explicitly and derive concrete directions for future work.
%
% Do not use the future-work chapter to hide limitations that affect the
% interpretation of the present results. Relevant limitations should also be
% discussed in the discussion chapter.
% ============================================================================

\cleardoublepage

\chapter{Limitations and Future Work}
\label{chap:future-work-limitations}

The findings of this thesis are subject to several limitations concerning the evaluated configuration space, the implemented selection procedures, and the available data basis. These limitations also motivate several directions for future research and system development.

\section{Limitations}
\label{sec:limitations}

The empirical findings are limited to the evaluated FINALLY version, the underlying dataset and algorithm snapshot, and the ten investigated configurations. Although the evaluation varied performance metrics, cutoffs, target-set sizes, filters, and required-dataset inclusion, it covered only part of the configuration space supported by the system. The results therefore establish constraint compliance, reproducibility, and strategy alignment only within the investigated conditions. They do not demonstrate that the generated dataset sets are scientifically suitable for every research question, produce more informative downstream evaluations, or provide practical advantages over manually constructed selections.

The selection procedures are additionally limited by the greedy search strategy and the preprocessing of the performance data. The greedy multi-start heuristic does not guarantee globally optimal dataset sets, and the evaluation establishes the direction of the resulting objective scores rather than their distance from the global optimum. FINALLY additionally applies mean imputation to missing performance values and does not standardize individual algorithm dimensions before selection. Missing dimensions are therefore moved toward the corresponding column mean, while algorithm dimensions with greater performance variation can contribute more strongly to Euclidean distances, covariance, and the resulting diversity scores. Alternative preprocessing decisions could consequently lead to different selected dataset sets.

The implementations developed for FINALLY adapt Reising's existing \gls{aps}-based diversity measures. The objectives use raw scores without set-size-specific normalization and omit the additive stabilizer \(\varepsilon\). The missing normalization limits comparisons of absolute score magnitudes across different configurations, while omission of \(\varepsilon\) causes the complete objective to become zero whenever its structural component is zero, independently of the corresponding \gls{mnnd}. These choices did not prevent complete directional alignment in the final evaluation, but they remain relevant when applying FINALLY to performance spaces with different geometric properties.

The corrected Convex Hull implementation produced positive, directionally aligned scores in all configurations of the final evaluation. However, the evaluation did not exercise every geometric path supported by the implementation. In particular, more general high-dimensional hull configurations containing more than the minimum number of points for their intrinsic rank may exhibit numerical conditions not represented by the evaluated dataset sets. A numerically induced zero hull volume would be particularly relevant for a minimizing strategy because it could be preferred over a genuinely compact set with a small but positive score. FINALLY can only recommend datasets represented by the available metadata and algorithm-performance data. Its applicability therefore depends on the coverage and quality of this data basis.

\section{Future Work}
\label{sec:future-work}

Future evaluations should test the numerical robustness and scalability of the \gls{aps}-based selection procedures beyond the geometric cases covered by the final evaluation. In particular, the general high-dimensional Convex Hull path should be tested systematically with point sets whose intrinsic rank is smaller than the number of points minus one, including cases with very small but valid hull volumes. Such tests could determine whether additional scale-aware numerical criteria are required for the facet-based hull construction. Recording intermediate candidate sets and objective values would also make the greedy search more traceable and simplify the diagnosis of numerical or optimization problems.

Scalability should also be investigated independently of the complete web workflow. The present end-to-end evaluation showed increasing response times as the requested target-set size grew, but these measurements combine selection computation with browser, network, server, and interface effects. Controlled algorithm-level benchmarks over larger candidate pools, target-set sizes, and algorithm dimensions could isolate the computational cost of Convex Hull and \gls{effcov} selection and identify the main scalability bottlenecks.

Further evaluations should cover additional dataset and algorithm snapshots, performance metrics, cutoffs, filter combinations, required datasets, candidate-pool sizes, and target-set sizes. Sensitivity analyses should examine how recommendations depend on mean imputation, alternative missing-value treatments, feature-wise standardization, algorithm-portfolio composition, numerical thresholds, and the number of greedy search initializations. These experiments would show whether the observed strategy alignment and selected dataset sets remain stable under alternative preprocessing decisions and performance-space representations.

Future work should investigate whether FINALLY-generated dataset sets lead to different or more informative experimental conclusions than established dataset-selection practices. Equal-sized sets generated through diverse, non-diverse, and Random selection could be compared with sets composed of frequently used benchmark datasets and with selections observed in published recommender-systems studies. The same algorithms and evaluation protocol should then be applied to all sets to determine whether they expose different performance patterns, algorithm rankings, strengths, or weaknesses. Such a downstream comparison would move beyond measuring alignment with FINALLY's own objectives and examine whether performance-oriented dataset-set selection provides additional empirical insight relative to commonly used alternatives.

Studies with recommender-systems researchers could compare manual dataset selection with selection using FINALLY and examine how users interpret the strategies, inspect recommendations, and justify their final decisions. The system could provide more detailed explanations of how individual datasets contribute to a set-level objective, expose relevant score and data-quality information, and support additional or multi-objective selection strategies. Further extensions could broaden the dataset and performance corpus, improve the integration of authoritative dataset citations, and introduce a configurable random seed for reproducible Random recommendations.

% ============================================================================
% DECLARATION ON THE USE OF GENERATIVE AI
% ============================================================================

\cleardoublepage

\phantomsection
\addcontentsline{toc}{chapter}{Generative AI Usage Disclosure}
\chapter*{Generative AI Usage Disclosure}

I used generative AI tools, including coding assistants, as supporting tools during the preparation of this thesis. They assisted with language refinement, structuring and formulation of explanations, the development and revision of source code, debugging, and the preparation of technical explanations and documentation. All research decisions, system requirements, experimental design, implementation choices, analyses, and interpretations presented in this thesis were critically reviewed and verified by myself. I remain fully responsible for the content and results of this work.

% ============================================================================
% BIBLIOGRAPHY
%
% Store the bibliography entries in references.bib.
% Every citation key used in the thesis must exist in that file.
% ============================================================================

\cleardoublepage

\phantomsection

\addcontentsline{toc}{chapter}{Bibliography}

\bibliography{references}

% ============================================================================
% APPENDICES
%
% Use appendices for material that is relevant but not essential to the main
% argument, for example source code, questionnaires, additional tables,
% detailed proofs, or complete experimental results.
% ============================================================================

\cleardoublepage

\appendix

\chapter{Supplementary Material}
\label{app:supplementary}

\section{Additional Interface Views}
\label{app:interface-views}

\subsection{Advanced Range Filters}

\begin{figure}[htbp]
    \centering
    \includegraphics[
        width=0.9\textwidth,
        height=0.75\textheight,
        keepaspectratio
    ]{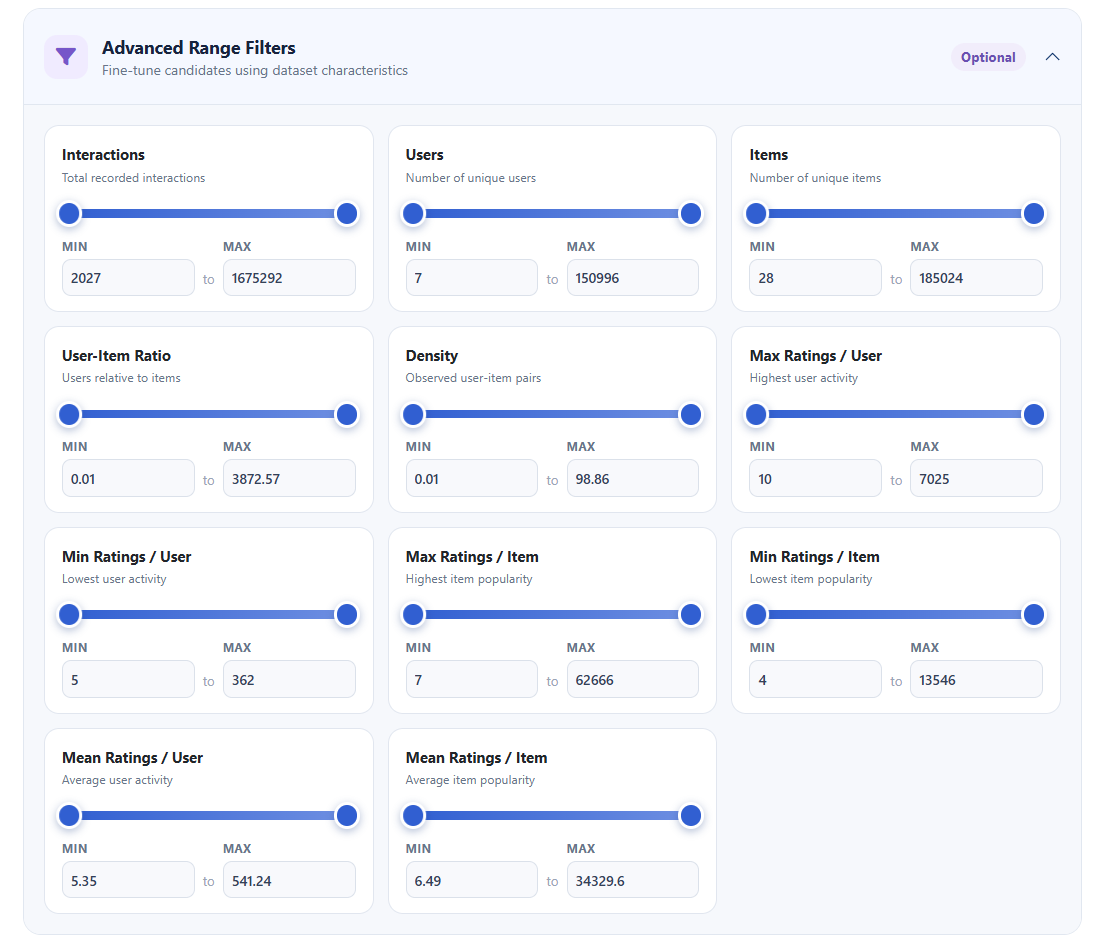}
    \caption{Expanded \emph{Advanced Range Filters} in Step~2 of FINALLY. The controls restrict the eligible candidate pool according to dataset characteristics before the recommendation is generated.}
    \label{fig:finally-range-filters}
\end{figure}

\subsection{Markdown Export Dialog}

\begin{figure}[htbp]
    \centering
    \includegraphics[
        width=0.9\textwidth,
        height=0.75\textheight,
        keepaspectratio
    ]{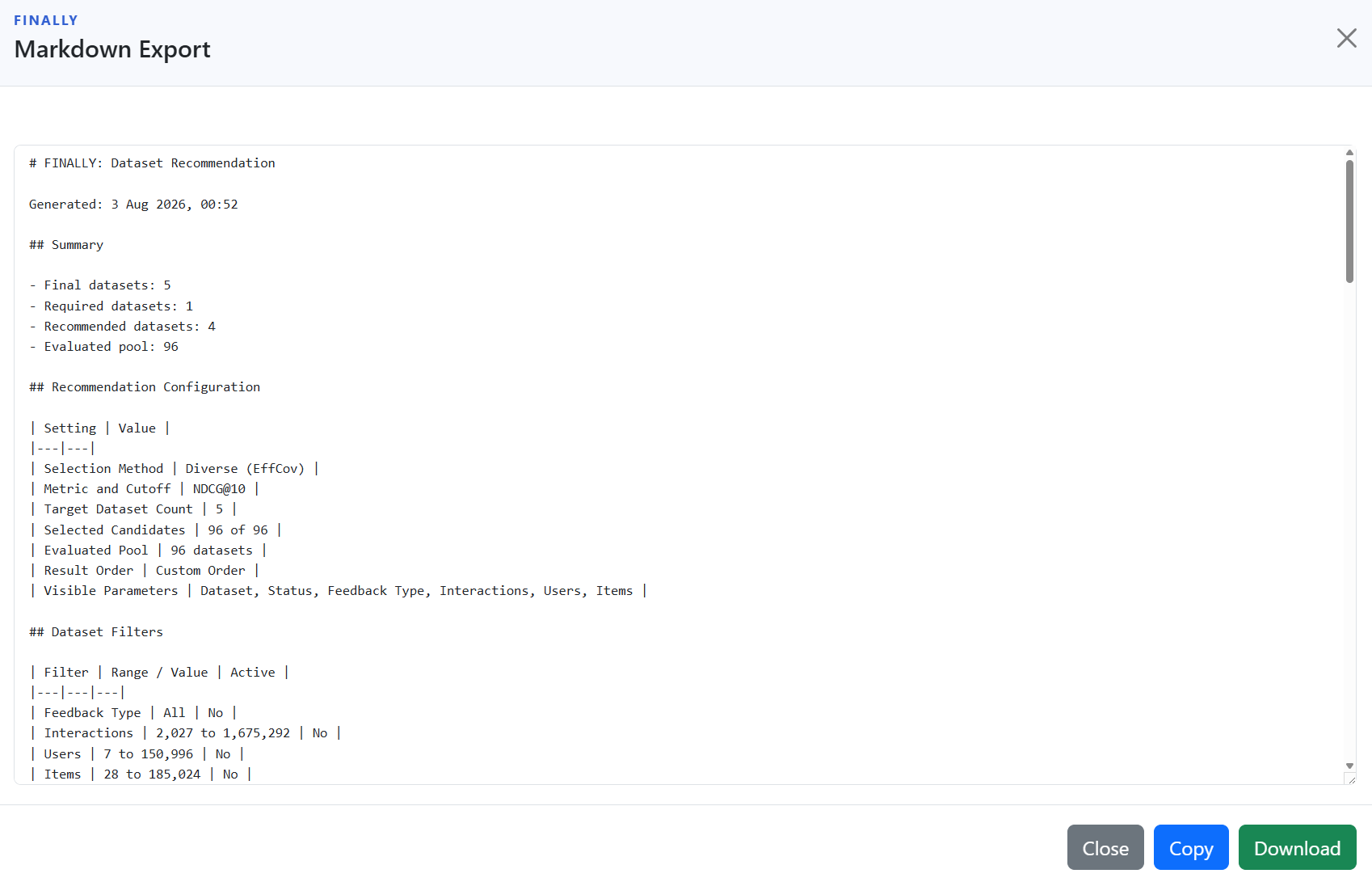}
    \caption{Markdown export dialog in FINALLY, including the generated source preview and the controls for copying and downloading the export.}
    \label{fig:finally-markdown-export}
\end{figure}

\section{Complete Response-Time Statistics}
\label{app:response-time-statistics}

Table~\ref{tab:complete-response-times} reports the complete descriptive end-to-end response-time statistics for all 50 configuration--strategy combinations evaluated in Chapter~\ref{chap:evaluation}. Times are reported in milliseconds because the browser-based collector recorded durations at millisecond resolution. The standard deviation is the population standard deviation. Deterministic strategies comprise three repetitions per configuration, whereas Random selection comprises 30 repetitions.

\begin{table}[p]
    \centering
    \scriptsize
    \setlength{\tabcolsep}{3pt}
    \renewcommand{\arraystretch}{0.86}

    \caption{Complete end-to-end response-time statistics for all evaluated configuration--strategy combinations.}
    \label{tab:complete-response-times}

    \begin{tabular}{@{}llrrrrr@{}}
        \toprule
        Configuration & Strategy & \(n\) & Mean & SD & Min. & Max. \\
        \midrule

        \texttt{ndcg10-size5}
        & D-CH  & 3  & 277.3 & 2.4 & 274 & 279 \\
        & ND-CH & 3  & 279.3 & 1.2 & 278 & 281 \\
        & D-EC  & 3  & 236.3 & 0.5 & 236 & 237 \\
        & ND-EC & 3  & 235.0 & 2.9 & 231 & 238 \\
        & R     & 30 & 152.6 & 9.3 & 139 & 169 \\
        \midrule

        \texttt{ndcg10-size10}
        & D-CH  & 3  & 1200.0 & 5.9 & 1194 & 1208 \\
        & ND-CH & 3  & 1214.3 & 3.9 & 1209 & 1218 \\
        & D-EC  & 3  & 997.3  & 4.6 & 991  & 1002 \\
        & ND-EC & 3  & 1009.7 & 0.9 & 1009 & 1011 \\
        & R     & 30 & 150.8  & 9.5 & 139  & 171 \\
        \midrule

        \texttt{ndcg10-size20}
        & D-CH  & 3  & 9433.7 & 35.8 & 9383 & 9459 \\
        & ND-CH & 3  & 9589.3 & 44.8 & 9550 & 9652 \\
        & D-EC  & 3  & 8325.7 & 34.8 & 8285 & 8370 \\
        & ND-EC & 3  & 8515.7 & 42.9 & 8480 & 8576 \\
        & R     & 30 & 151.0  & 8.1  & 137  & 166 \\
        \midrule

        \texttt{hr10-size5}
        & D-CH  & 3  & 276.0 & 1.4 & 275 & 278 \\
        & ND-CH & 3  & 279.0 & 4.3 & 273 & 283 \\
        & D-EC  & 3  & 235.3 & 3.4 & 232 & 240 \\
        & ND-EC & 3  & 239.3 & 0.5 & 239 & 240 \\
        & R     & 30 & 149.6 & 9.0 & 139 & 171 \\
        \midrule

        \texttt{recall10-size5}
        & D-CH  & 3  & 277.7 & 5.3 & 271 & 284 \\
        & ND-CH & 3  & 278.3 & 3.3 & 274 & 282 \\
        & D-EC  & 3  & 235.7 & 2.1 & 233 & 238 \\
        & ND-EC & 3  & 241.3 & 5.4 & 237 & 249 \\
        & R     & 30 & 148.1 & 7.7 & 139 & 164 \\
        \midrule

        \texttt{ndcg1-size5}
        & D-CH  & 3  & 277.7 & 0.5 & 277 & 278 \\
        & ND-CH & 3  & 279.0 & 2.4 & 276 & 282 \\
        & D-EC  & 3  & 234.7 & 0.9 & 234 & 236 \\
        & ND-EC & 3  & 237.0 & 2.2 & 235 & 240 \\
        & R     & 30 & 152.0 & 7.6 & 140 & 165 \\
        \midrule

        \texttt{ndcg20-size5}
        & D-CH  & 3  & 275.7 & 2.1 & 273 & 278 \\
        & ND-CH & 3  & 280.3 & 1.7 & 278 & 282 \\
        & D-EC  & 3  & 238.0 & 2.2 & 236 & 241 \\
        & ND-EC & 3  & 237.3 & 2.6 & 235 & 241 \\
        & R     & 30 & 152.3 & 8.9 & 141 & 171 \\
        \midrule

        \texttt{filtered-ndcg10-size5}
        & D-CH  & 3  & 494.0 & 7.9 & 487 & 505 \\
        & ND-CH & 3  & 499.0 & 5.7 & 491 & 503 \\
        & D-EC  & 3  & 492.7 & 1.2 & 491 & 494 \\
        & ND-EC & 3  & 493.0 & 1.6 & 491 & 495 \\
        & R     & 30 & 490.5 & 9.6 & 469 & 511 \\
        \midrule

        \texttt{required-ndcg10-size5}
        & D-CH  & 3  & 147.3 & 0.5 & 147 & 148 \\
        & ND-CH & 3  & 146.3 & 5.7 & 139 & 153 \\
        & D-EC  & 3  & 148.0 & 3.6 & 145 & 153 \\
        & ND-EC & 3  & 142.7 & 1.9 & 140 & 144 \\
        & R     & 30 & 146.1 & 5.5 & 138 & 156 \\
        \midrule

        \texttt{filtered-required-ndcg10-size5}
        & D-CH  & 3  & 491.3 & 4.1 & 486 & 496 \\
        & ND-CH & 3  & 487.7 & 6.0 & 482 & 496 \\
        & D-EC  & 3  & 488.0 & 5.7 & 481 & 495 \\
        & ND-EC & 3  & 484.0 & 9.4 & 475 & 497 \\
        & R     & 30 & 490.7 & 7.5 & 474 & 509 \\

        \bottomrule
    \end{tabular}

    \vspace{0.3em}

    \parbox{\textwidth}{\scriptsize
        \textit{Notes:} Times are reported in milliseconds.
        D-CH = Diverse Convex Hull; ND-CH = Non-Diverse Convex Hull;
        D-EC = Diverse EffCov; ND-EC = Non-Diverse EffCov;
        R = Random. SD denotes the population standard deviation.
    }
\end{table}

\cleardoublepage

% ============================================================================
% DECLARATION OF AUTHORSHIP
%
% This declaration is a separate chapter in the back matter and appears in
% the table of contents.
%
% Check whether the applicable examination regulations require an additional
% institution-specific form, handwritten signature, place, or date.
% ============================================================================

\backmatter

\chapter{Declaration of Authorship}

I certify that I have prepared this written work independently and have not used any aids other than those specified. I have clearly identified all passages that are taken from other works in terms of wording or meaning (including translations) as borrowed material in each individual case, stating the exact source (including the World Wide Web as well as generative AI and other electronic data collections). This also applies to attached drawings, pictorial representations, sketches and the like. I acknowledge that the proven omission of the indication of origin will be regarded as attempted deception. I confirm that the content of the electronic version is the same as the printed version.

\vspace{2cm}

\noindent
\begin{tabularx}{\textwidth}{@{}X X@{}}

    \rule{0.42\textwidth}{0.4pt}
    &
    \rule{0.42\textwidth}{0.4pt}
    \\

    Place and date
    &
    Signature

\end{tabularx}

\thispagestyle{empty} %edited
\end{document}